\documentclass[11pt]{article}
\usepackage{latexsym}

\usepackage{epsfig}
\usepackage{graphicx}

\catcode`\@=11

\@addtoreset{equation}{section}

\def\@normalsize{\@setsize\normalsize{15pt}\xiipt\@xiipt
\abovedisplayskip 14pt plus3pt minus3pt%
\belowdisplayskip \abovedisplayskip
\abovedisplayshortskip  \z@ plus3pt%
\belowdisplayshortskip  7pt plus3.5pt minus0pt}
\def\small{\@setsize\small{13.6pt}\xipt\@xipt
\abovedisplayskip 13pt plus3pt minus3pt%
\belowdisplayskip \abovedisplayskip
\abovedisplayshortskip  \z@ plus3pt%
\belowdisplayshortskip  7pt plus3.5pt minus0pt
\def\@listi{\parsep 4.5pt plus 2pt minus 1pt
            \itemsep \parsep
            \topsep 9pt plus 3pt minus 3pt}}

\def\underline#1{\relax\ifmmode\@@underline#1\else
        $\@@underline{\hbox{#1}}$\relax\fi}
\@twosidetrue \relax

\catcode`@=12

\catcode`\@=11

\def\section{\@startsection{section}{1}{\z@}{3.5ex plus 1ex minus
   .2ex}{2.3ex plus .2ex}{\large\bf}}

\def\ps@headings{\def\@oddfoot{}\def\@evenfoot{}
\def\@oddhead{\hbox{}\hfill
        \makebox[.5\textwidth]{\raggedright\ignorespaces --\thepage{}--
        \hfill }}
\def\@evenhead{\@oddhead}
\def\subsectionmark##1{\markboth{##1}{}}
}

\ps@headings

\catcode`\@=12

\newcommand{\be}{\begin{equation}}
\newcommand{\ee}{\end{equation}}
\newcommand{\bea}{\begin{eqnarray}}
\newcommand{\nn}{\nonumber}
\newcommand{\eea}{\end{eqnarray}}

\begin{document}

\begin{titlepage}
\begin{flushright}
gr-qc/0608026\\ August 2006
\end{flushright}

\vspace{0.30in}
\begin{centering}

{\large {\bf Curvaton Dynamics in Brane-worlds}}
\\

\vspace{0.7in} {\bf  Eleftherios Papantonopoulos$^{*}$ and
Vassilios Zamarias$^\flat$}

\vspace{0.04in}

Department of Physics, National Technical University of Athens,\\
Zografou Campus GR 157 73, Athens, Greece\\
\end{centering}

\vspace{0.8in}

\begin{abstract}
We study the curvaton dynamics in brane-world cosmologies.
Assuming that the inflaton field survives without decay after the
end of inflation, we apply the curvaton reheating mechanism to
Randall-Sundrum and to its curvature corrections: Gauss-Bonnet,
induced gravity and combined Gauss-Bonnet and induced gravity
cosmological models. In the case of chaotic inflation and
requiring suppression of possible short-wavelength generated
gravitational waves, we constraint the parameters of a successful
curvaton brane-world cosmological model. If density perturbations
are also generated by the curvaton field then, the fundamental
five-dimensional mass could be much lower than the Planck mass.

 \end{abstract}

\begin{flushleft}

\vspace{0.9in} $^{*}$lpapa@central.ntua.gr \\
$^{\flat}$zamarias@central.ntua.gr
\end{flushleft}
\end{titlepage}

\section{Introduction}

Inflation is an indispensable part of the hot big bang cosmology,
solving long standing problems of standard cosmology, such as
homogeneity, isotropy and flatness of the universe
\cite{inflation}. Moreover, it generates superhorizon fluctuations
which become classical after crossing out the Hubble horizon
 and seed the matter and radiation fluctuations observed
in the universe after re-entry of the horizon  at the end of the
inflation. The microwave anisotropy encodes information from this
early inflationary phase of the cosmological evolution. The
three-year WMAP result as well as other astronomical data
\cite{3jwmap} supports the inflation scenario of the hot big bang
cosmology (for a review see \cite{lazarides}).

The results of WMAP three-year data presented in \cite{3jwmap} do
not favour a scale-invariant spectrum of fluctuations giving the
value of the index of the power spectrum as
$n_s=0.951^{+0.015}_{-0.019}.$ Also, these results favour the
simple chaotic inflation model with potential $m^2\phi^2$ which
fits the observations very well. The WMAP three-year data provides
then significant constraints on the inflation models and has ruled
out some of them \cite{alabidi}.

In spite of many phenomenological successes of inflation, there
nevertheless many serious problems remain to be understood, such
as the initial cosmological singularity problem, the
trans-Planckian problem, and problems concerning the very
existence of the scalar field, the inflaton, driving inflation.
The general believe is that these problems can only be addressed
in a more general framework than effective field theory, such as
string theory the only self-consistent theory till now.

Recently  effort was spent in addressing the problem of inflation
in brane-world models. The most successful model that incorporates
the idea that our universe lies in a three-dimensional brane
within a higher-dimensional bulk spacetime is the Randall-Sundrum
model of a single brane in an AdS bulk \cite{randall}. There are
also other brane cosmological models which are merely curvature
generalizations of the Randall-Sundrum model and give novel
features compared to standard cosmology. The induced gravity
cosmological model \cite{dvali1} arises when we add to the brane
action a four-dimensional scalar curvature term, while the
Gauss-Bonnet model \cite{germani} arises when we include a
Gauss-Bonnet term to the five-dimensional action. Finally, if both
terms are included in the action, the combined cosmological model
\cite{papa} describes their cosmological evolution (for reviews on
the brane-world cosmological models see \cite{reviews}).

All these brane-world inflationary models have in the high energy
limit correction terms in their Friedmann equations. These terms
have important consequences in the inflationary dynamics. In the
case of steep inflation~\cite{cope}, the exponential potential is
so steep that it would not drive inflation in the standard
cosmology, but it may do so in the presence of these corrections
terms.  In light of the WMAP three-year data, the chaotic
inflation braneworld models with simple quadratic inflaton
potentials are more favoured, giving a value for the index of the
power spectrum, close to observations~\cite{MaartensChaoticRS2}.
In these models the correction terms assist inflation damping the
kinetic energy of the inflaton field. However, as the energy
density decreases, these corrections become unimportant, and the
inflaton field enters a kinetic energy dominated regime, bringing
inflation to an end. As the inflaton may survive this process
without decay, an alternative reheating mechanism is required.

The curvaton reheating mechanism~\cite{feng} was proposed as an
alternative mechanism to complement the other two known
mechanisms, the conventional decay of the inflaton energy density
into ordinary matter~\cite{Bassett:2005xm} and the mechanism of
gravitational particle production at the end of
inflation~\cite{gravpart}. The curvaton scenario was firstly
 suggested as an alternative mechanism to generate the primordial
 scalar perturbation which is responsible for the structure formation.
In this scenario the primordial density perturbation originates
from the vacuum
 fluctuation of some ``curvaton'' field $\sigma$, different from the
  inflaton field \cite{curvaton}. In brane-worlds the curvaton reheating mechanism
was employed in~\cite{liddle}, where it was shown that it can
overcome the problems encountered in the reheating process  via
gravitational particle production \cite{cope,gravpart,MJ1} in
steep inflation, allowing a high reheat temperature and preventing
short-wavelength gravitational wave dominance \cite{sahni}.

If the inflaton field is to survive without decay at the end of
inflation, then it may play the r$\hat{o}$le of the quintessence
field~\cite{Peebles:1998qn}. In this case the inflaton  field
enters a long kinetic epoch and an alternative mechanism for the
reheating is required. Thus in all Quintessential Inflation
models~\cite{quintessence_models} the need of the curvaton field
proves to be essential. Moreover, the Quintessential Inflation can
also be applied to brane-worlds~\cite{Brane_quintessence_models}.

 In this work we will study in a
systematic way the curvaton dynamics in brane-world cosmological
models. The curvaton dynamics can be introduced in brane-worlds in
various ways. In their high energy limit, depending on the
parameters of the model, we have transitions, as the energy
density decreases, from one dimensionality spacetime to another of
different dimensionality. If inflation occurs in one spacetime and
survives without decay the curvaton reheating can occur in the
same dimensionality spacetime or in a different dimensionality
one. For example, in the induced gravity model there is a choice
of parameters for which the universe can start in four dimensions
pass from a five-dimensional phase and then again end up in four
dimensions before
nucleosynthesis~\cite{lpapzamchaotic,Zhang:2004in}.

From the recent observation data there is no any compelling reason
to decide whether the primordial density perturbations originate
from the vacuum fluctuations of an inflaton or curvaton field. If
the curvaton field is employed for both reheating and primordial
density perturbations, since it is not directly linked with the
high energy scales of the inflaton field, it can be associated
with lower energy particles of the TeV region~\cite{lowcurvaton}.
We will show that in brane-worlds if the density perturbations are
generated by the curvaton field then the energy scale of the
fifth-dimension can be much below the $M_{Pl}$ scale.

Another important constraint that all the curvaton models should
satisfy in the case the inflaton survives without decay, is that
short-wavelength gravitational waves generated during the kinetic
epoch should not dominate over radiation. In all cases we
consider, we will derive constraints which the parameters should
satisfy in order to suppress the gravitational radiation dominance
during the kinetic period.

To capture the curvaton dynamics, we will develop a general
curvaton formalism for two
 different cosmological regimes followed one another and which are
 characterized by different Friedmann
equations, corresponding to two different dimensionality
spacetimes in brane-world cosmologies. Constraints on the curvaton
parameters in these two regimes will be derived and also
constrained relations on the parameters will be extracted from the
requirement of not having gravitational waves dominance. Then,
this general formalism for quadratic potentials for both the
inflaton and curvaton field will be applied to Randall-Sundrum,
Gauss-Bonnet, induced gravity and combined Gauss-Bonnet and
induced gravity
 cosmological models. We will get bounds on the
various parameters first with the assumption that the inflaton
field is responsible for the primordial density perturbations and
the curvaton field for the reheating and second with the
assumption that both density perturbations and reheating are
generated by the curvaton field.

\section{General Curvaton Formalism}

We assume that the cosmological evolution of the universe is
described by two different cosmological regimes characterized by
the following general Friedmann equations \bea
H_{1}^{2} &=& \beta_{1}\, \rho_{1}^{\alpha_{1}}, \label{Friedmann1}\\
H_{2}^{2} &=& \beta_{2}\, \rho_{2}^{\alpha_{2}},
\label{Friedmann2} \eea where $\beta_{1}$ and $\beta_{2}$ are
proportional to the Newton's constants in the two regimes and the
powers $\alpha_{1}$ and $\alpha_{2}$ may take values that they
give conventional or unconventional energy densities.  The
transition from the one regime to the other occurs at \bea
\rho_{1.2} &=& \Big{(} \frac{\beta_{2}}{\beta{1}} \Big{)}
^{\frac{1}{\alpha_{1}-\alpha_{2}}}, \label{rhotransit}\\
H_{1.2} &=& \beta_{2}^{1/2}\, \Big{(} \frac{\beta_{2}}{\beta{1}}
\Big{)} ^ {\frac{\alpha_{2}}{2(\alpha_{1}-\alpha_{2})}}.
\label{Friedtransit} \eea We consider on the brane two scalar
fields, the inflaton field $\phi$ and a curvaton field $\sigma$
with no interactions between them. Their energy densities are
given by \bea
\rho_{\phi}= \frac{\dot{\phi}^{2}}{2} + V(\phi)~, \label{inflaton}\\
\rho_{\sigma}= \frac{\dot{\sigma}^{2}}{2} + U(\sigma)~,
\label{curvaton} \eea with their potentials of the form  \bea
V(\phi)&=&\delta\, \phi^{\gamma}, \\
U(\sigma)&=&\frac{1}{2}\, m^{2}\, \sigma^{2}. \eea The equations
of motion of the inflaton and curvaton fields are
 \bea
\ddot{\phi} + 3H\dot{\phi} + V'(\phi) = 0 \label{inflmotion}~,\\
\ddot{\sigma} + 3H\dot{\sigma} + U'(\sigma) = 0
\label{curvmotion}~. \eea

Providing that the curvaton field is responsible for the reheating
of the universe, the curvaton reheating mechanism can proceed  in three different ways: \\
\underline{1st Case}: The curvaton oscillates and decays into
radiation during
the regime $1$. Nucleosynthesis takes place during the regime $2$.\\
\underline{2nd Case}: The curvaton oscillates during the regime
$1$ but decays
into radiation in the regime $2$. \\
\underline{3rd Case}: The curvaton oscillation and decay take place during the regime $2$. \\
Depending on whether the curvaton becomes the dominant component
before or after decay we will have to consider two subcases for
each of the above three cases \bea
1a. \hspace{0.5in} m>H_{eq1}>\Gamma>H_{1.2}>H_{nucl} \nn \\
1b. \hspace{0.5in} m>\Gamma>H_{eq1}>H_{1.2}>H_{nucl} \nn \\
2a. \hspace{0.5in} m>H_{1.2}>H_{eq2}>\Gamma>H_{nucl} \nn \\
2b. \hspace{0.5in} m>H_{1.2}>\Gamma>H_{eq2}>H_{nucl} \nn \\
3a. \hspace{0.5in} H_{1.2}>m>H_{eq2}>\Gamma>H_{nucl} \nn \\
3b. \hspace{0.5in} H_{1.2}>m>\Gamma>H_{eq2}>H_{nucl} \nn
 \eea
where $m$ denotes the mass of the curvaton at the moment when it
starts to oscillate, $H_{eq1}$ and $H_{eq2}$ are the Hubble
parameters at the moment when the curvaton field or its decay
products starts to dominate over the inflaton field during the
regime $1$ or $2$ respectively,
 $\Gamma$ is the decay parameter of the curvaton field and $H_{nucl}$ the Hubble parameter
 at the moment when
nucleosynthesis starts. We have to consider two subcases depending
on whether the curvaton field comes to dominate in the universe
before or after it decays as radiation.  In both subcases the
crucial moment is when the energy density of the inflaton field
becomes equal to the energy density of the curvaton field
$\rho_{\sigma} = \rho_{\phi}$ at the moment of curvaton domination
$a=a_{eq1,2}$. In the following we will distinguish these two
different subcases and we will find constrained relations between
the three parameters $m,\Gamma,\sigma_{i}$, with $\sigma_{i}$ the
initial value of the curvaton field, for each of the above cases
separately.

\subsection{Curvaton Domination Before Decay}

\subsubsection{\underline{1st Case}: Oscillation and Decay in the
First Regime}

After the inflationary period begins the kinetic period (labeled
by 'kin') where the universe is still dominated by the inflaton
field where $\frac{1}{2}\, \dot{\phi}^{2} \gg V(\phi)$ and behaves
as stiff matter $(w=1)$ $\rho_{\phi} \propto a^{-6}$. Therefore
the energy density of the inflaton field evolves as \be
\rho_{\phi}=\rho_{\phi}^{(kin)}\, \Big{(} \frac{a_{kin}}{a}
\Big{)}^6, \ee and using (\ref{Friedmann1}) the Hubble parameter
evolves as \be H=H_{kin}\, \Big{(} \frac{a_{kin}}{a} \Big{)}
^{3\alpha_{1}}, \label{H1kin} \ee where \be
H_{kin}^{2}=\beta_{1}\, \rho_{\phi}^{(kin)\, \alpha_{1}}.
\label{Hkin} \ee During this stage the curvaton field is
effectively a constant keeping its initial value $\sigma_{i}$.

At some later time the curvaton field starts to oscillate (labeled
by 'osc'). Using (\ref{curvmotion}) we can see that this happens
when $H \simeq m$ while in this case the universe still evolves in
the regime $1$. In order to avoid a stage of curvaton driven
inflation the universe should still be dominated by the inflaton
field which means $\rho_{\sigma}^{(osc)} \ll \rho_{\phi}^{(osc)}$.
At this moment we have  \bea
\rho_{\sigma}^{(osc)} &=& \frac{1}{2}\, m^2\, \sigma_{i}^{2}~, \\
\rho_{\phi}^{(osc)} &=& \beta_{1}^{-1/\alpha_{1}}\,
m^{2/\alpha_{1}}, \eea which implies the following constraint on
the initial value of the curvaton field \be \sigma_{i}^{2} \ll 2\,
\beta_{1}^{-1/\alpha_{1}}\, m^{(2-2\alpha_{1})/\alpha_{1}}.
\label{curvinit} \ee The inflaton still decays as stiff matter \be
\rho_{\phi}=\rho_{\phi}^{(kin)}\, \frac{a_{kin}^6}{a_{osc}^6}\,
\frac{a_{osc}^6}{a^6}~, \label{infl1} \ee and using
(\ref{Friedmann1}) the Hubble parameter evolves as \be H=H_{kin}\,
\frac{a_{kin}^{3\alpha_{1}}}{a_{osc}^{3\alpha_{1}}}\,
\frac{a_{osc}^{3\alpha_{1}}}{a^{3\alpha_{1}}}~, \label{H1} \ee
where $H_{kin}^{2}$ is given by (\ref{Hkin}). The energy of the
curvaton field decays as non-relativistic matter $(w=0)$ \be
\rho_{\sigma}=\rho_{\sigma_{i}}\, \frac{a_{osc}^3}{a^3}~.
\label{curv1} \ee
 Then
using (\ref{infl1}), (\ref{curv1}) and (\ref{Hkin}) we find that
\be
\frac{1}{2}\,m^2\,\sigma_{i}^{2}\,\beta_{1}^{1/\alpha_{1}}\,H_{kin}^{-2/\alpha_{1}}\,
\frac{a_{osc}^{3}}{a_{kin}^{3}}\, =
\frac{a_{kin}^{3}}{a_{eq1}^{3}}~. \ee Moreover from (\ref{H1kin})
we have \be \frac{a_{osc}^{3}}{a_{kin}^{3}} =
\frac{H_{kin}^{1/\alpha_{1}}}{m^{1/\alpha_{1}}}~,
\ee obtaining  \be
\frac{1}{2}\,m^{2-1/\alpha_{1}}\,\sigma_{i}^{2}\,\beta_{1}^{1/\alpha_{1}}\,H_{kin}^{-1/\alpha_{1}}\,
= \frac{a_{kin}^{3}}{a_{eq1}^{3}}~. \label{key1} \ee From
(\ref{H1}) we finally conclude that \be H_{eq1} =
\frac{\beta_{1}}{2^{\alpha_{1}}}\, \sigma_{i}^{2\alpha_{1}}\,
m^{2\alpha_{1}-1}. \ee We can easily see that if in the regime $1$
we have the conventional four dimensional cosmological evolution,
which corresponds  to ${\alpha_{1}}=1$ and
$\beta_{1}=1/3M_{Pl}^{2}$, we find \be
H_{eq1}=\frac{\sigma_{i}^{2}}{6\,M_{Pl}^{2}}\,m~. \ee

\subsubsection{\underline{2nd Case}: Oscillation in the First
Regime and Decay in the Second Regime}

In this case, during the oscillation of the curvaton field the
universe undergoes a transition period from the regime $1$ to the
regime $2$. Therefore the energy density of the inflaton field
evolves as \be \rho_{\phi}=\rho_{\phi}^{(kin)}\,
\frac{a_{kin}^6}{a_{osc}^6}\, \frac{a_{osc}^6}{a_{1.2}^6}\,
\frac{a_{1.2}^6}{a^6}~, \label{infl2} \ee and using
(\ref{Friedmann1}) and (\ref{Friedmann2}) the Hubble parameter
evolves as \be H=H_{kin}\,
\frac{a_{kin}^{3\alpha_{1}}}{a_{osc}^{3\alpha_{1}}}\,
\frac{a_{osc}^{3\alpha_{1}}}{a_{1.2}^{3\alpha_{1}}}\,
\frac{a_{1.2}^{3\alpha_{2}}}{a^{3\alpha_{2}}}~, \label{H2} \ee
where (\ref{Hkin}) still holds. The energy of the curvaton field
decays as non-relativistic matter $(w=0)$ \be
\rho_{\sigma}=\rho_{\sigma_{i}}\, \frac{a_{osc}^3}{a_{1.2}^3}\,
\frac{a_{1.2}^3}{a^3}~. \label{curv2} \ee

Then for subcase $2a.$ we need the value of $H_{eq2}$ when
$\rho_{\sigma} = \rho_{\phi}$ at $a=a_{eq2}$. Here we can rewrite
equation (\ref{H2}) as \be H=H_{kin}\,
\frac{a_{kin}^{3(\alpha_{1}-\alpha_{2})}}{a_{1.2}^{3(\alpha_{1}-\alpha_{2})}}\,
\frac{a_{kin}^{3\alpha_{2}}}{a_{eq2}^{3\alpha_{2}}}~. \label{H2'}
\ee Using (\ref{infl2}), (\ref{curv2}) and (\ref{Hkin}) we find as
in case $1a.$ that \be
\frac{1}{2}\,m^{2-1/\alpha_{1}}\,\sigma_{i}^{2}\,\beta_{1}^{1/\alpha_{1}}\,H_{kin}^{-1/\alpha_{1}}\,
= \frac{a_{kin}^{3}}{a_{eq2}^{3}}~. \label{key2} \ee Moreover from
(\ref{H2}) we have \be \frac{a_{kin}^{3}}{a_{1.2}^{3}} =
\frac{H_{1.2}^{1/\alpha_{1}}}{H_{kin}^{1/\alpha_{1}}}~, \ee and we
find that \be H_{eq2} =
\frac{\sigma_{i}^{2\alpha_{2}}}{2^{\alpha_{2}}}\,
m^{2\alpha_{2}-\frac{\alpha_{2}}{\alpha_{1}}} \,
\beta_{1}^{\frac{\alpha_{2}}{2{\alpha_{1}}}}\, \beta_{2}^{1/2}~.
\ee

\subsubsection{\underline{3rd Case}: Oscillation and Decay in the
Second Regime}

The results are the same as in the first case replacing all
indices $1$ by $2$. Thus we have \be H_{eq2} =
\frac{\beta_{2}}{2^{\alpha_{2}}}\, \sigma_{i}^{2\alpha_{2}}\,
m^{2\alpha_{2}-1}. \ee

\subsection{Curvaton Decay Before  Domination}

\subsubsection{\underline{1st Case}: Oscillation and Decay in the
First Regime}

In this section $2.2$ we consider the case where the curvaton
field starts decaying as radiation before it comes to dominate the
inflaton field. Therefore the universe is still dominated by the
inflaton field and for subcase $1b.$ its energy density evolves as
\be \rho_{\phi}=\rho_{\phi}^{(kin)}\,
\frac{a_{kin}^6}{a_{osc}^6}\, \frac{a_{osc}^6}{a^6}~,
\label{infl1b} \ee and using (\ref{Friedmann1}) the Hubble
parameter evolves as \be H=H_{kin}\,
\frac{a_{kin}^{3\alpha_{1}}}{a_{osc}^{3\alpha_{1}}}\,
\frac{a_{osc}^{3\alpha_{1}}}{a_{d}^{3\alpha_{1}}}\,
\frac{a_{d}^{3\alpha_{1}}}{a^{3\alpha_{1}}}~, \label{H1b} \ee
where 'd' labels the quantities at the time of curvaton decay.
Here (\ref{Hkin}) still holds. The energy of the curvaton field
decays as radiation $(w=1/3)$ \be
\rho_{\sigma}=\rho_{\sigma_{i}}\, \frac{a_{osc}^3}{a_{d}^3}\,
\frac{a_{d}^4}{a^4}~. \label{curv1b} \ee At decay ($H=\Gamma$) we
have \be \label{decay1b} \frac{\Gamma}{H_{kin}} =
\frac{a^{3\alpha_{1}}_{kin}}{a^{3\alpha_{1}}_{d}}~, \ee and using
(\ref{infl1b}), (\ref{curv1b}) and (\ref{Hkin}) we find that \be
\frac{1}{2}\,m^{2-1/\alpha_1}\,\sigma_{i}^{2}\,\beta_{1}^{1/\alpha_{1}}\,H_{kin}^{-2/3\alpha_{1}}\,
\Gamma^{-1/3\alpha_1}\, = \frac{a_{kin}^{2}}{a_{eq1}^{2}}~. \ee
Then from (\ref{H1b}) we finally conclude that \be H_{eq1} =
\Big{(}\frac{\beta_{1}}{2^{\alpha_{1}}}\Big{)}^{3/2}\,
\sigma_{i}^{3\alpha_{1}}\, m^{3\alpha_{1}-3/2}\,\Gamma^{-1/2}. \ee

\subsubsection{\underline{2nd Case}: Oscillation in the First
Regime and Decay in the Second Regime}

In this case, one has to pay attention that at decay relation
(\ref{decay1b}) does not hold but \be \label{decay2b}
\frac{a_d}{a_{kin}} =
\frac{\beta_2^{1/6\alpha_2}\,H_{kin}^{1/3\alpha_1}}
{\beta_1^{1/6\alpha_1}\,\Gamma^{1/3\alpha_2}}~. \ee As previously
we obtain \be
\frac{1}{2}\,m^{2-1/\alpha_1}\,\sigma_{i}^{2}\,\beta_{1}^{5/6\alpha_{1}}\,H_{kin}^{-2/3\alpha_{1}}\,
\beta_{2}^{1/6\alpha_2}\,\Gamma^{-1/3\alpha_2}\, =
\frac{a_{kin}^{2}}{a_{eq1}^{2}}~, \ee and finally \be H_{eq2} =
\frac{\sigma_{i}^{3\alpha_{2}}\,m^{3\alpha_{2}-3\alpha_2/2\alpha_1}}{2^{3\alpha_2/2}}
\,\beta_{1}^{3\alpha_2/4\alpha_1}\,\beta_2^{3/4}\,\Gamma^{-1/2}.
\ee

\subsubsection{\underline{3rd Case}: Oscillation and Decay in the
Second Regime}

The results are the same as in the first case replacing all
indices $1$ by $2$. Thus we have
\be
H_{eq2} =
\Big{(}\frac{\beta_{2}}{2^{\alpha_{2}}}\Big{)}^{3/2}\, \sigma_{i}^{3\alpha_{2}}\,
m^{3\alpha_{2}-3/2}\,\Gamma^{-1/2}. \ee

\subsection{Gravitational Waves and Curvaton Dynamics}

Gravitational waves behave as massless scalar fields and their
amplitude remains constant during inflation, therefore we can
identify the amplitude of the gravitational waves $h^{2}_{GW}$
with the amplitude of the scalar tensor perturbations $A^{2}_{T}$
in the weak field approximation. During the kinetic epoch, the
background is dominated by the inflaton field, thus the energy
density of the background is the one of the inflaton scalar field
which decays as stiff matter $(\propto a^{-6})$ and the
gravitational waves $(\propto a^{-4})$ evolves as \footnote{Note
that the formula~(\ref{GWkin}) describes the gravitational
radiation in four-dimensional gravity. In brane-worlds the brane
is embedded in five-dimensional spacetime. However, we can use the
above formula because as it was shown in~\cite{Garriga:1999yh} the
gravitational waves are localized on the world volume of the
four-dimensional brane and the five-dimensional effects are
negligible.} \be \rho_{g}=\frac{32}{3\pi}\, h^{2}_{GW}\,
\rho_{\phi}\, \Big{(} \frac{a}{a_{kin}} \Big{)}^2. \label{GWkin}
\ee

We will discuss first the case of curvaton domination before it
decays. At the epoch of stiff scalar matter equality with curvaton
matter $(\rho_{\sigma}=\rho_{\phi})$, we have $\rho_{g} \ll
\rho_{\sigma}$. From equation (\ref{GWkin}) with the use of
equation (\ref{key1}) for cases $1a., 2a., 3a.$, we obtain the
following constraint \be \frac{\rho_{g}}{\rho_{\sigma}}|_{a=a_{eq}}=\frac{32}{3\pi}\,
h^{2}_{GW}\, \Big{(} \frac{2\, H_{kin}^{1/\alpha_{1}}}
{m^{2-1/\alpha_{1}}\, \beta_{1}^{1/\alpha_{1}}\, \sigma_{i}^{2}}
\Big{)} ^{2/3} \ll 1~,  \label{GWeq} \ee where for the
case $3a.$ the replacement of indices $1$ with $2$ is understood.
Then the curvaton decays before nucleosynthesis and we have the
following constraint for the decay parameter $\Gamma$ \be
H_{nucl}=10^{-40}\, M_{Pl} < \Gamma < H_{eq}~. \ee

Next we consider the case of the decay of the curvaton before it
dominates the universe. We will first discuss the case $2b.$ where
kination and oscillation occur during the 1 regime while the
curvaton decays during the 2 regime. The curvaton field decays
producing radiation at the time when $\Gamma=H$ and we have (\ref{decay2b}).
The produced radiation  evolves  as \be
\rho_{rad}^{(\sigma)}=\rho_{\sigma_{i}}\,
\frac{a^{3}_{osc}}{a^{3}_d}\, \frac{a^{4}_{d}}{a^{4}}~. \ee Then,
using the expression (\ref{infl2}) of the evolution of the
inflaton field we find that \be
\frac{a_{kin}^{2}}{a_{d}^{2}}=\frac{1}{2}\, m^{2-1/\alpha_{1}}\,
\sigma_{i}^{2}\, \beta_{1}^{5/6\alpha_{1}}\,
H_{kin}^{-2/3\alpha_{1}}\, \Gamma^{-1/3\alpha_{2}}\,\beta_{2}^{1/6\alpha_{2}}. \label{keyGW}
\ee The radiative regime is reached when $a=a_{d}$ where the
energy density of the background is $
 \rho_{\phi} +
\rho_{rad} \simeq 2 \rho_{rad}$. We used $\rho_{rad}$ to point out
that during the decay phase the curvaton decays as radiation. Then
the evolution of the gravitational waves is given by \be
\rho_{g}=\frac{64}{3\pi}\, h^{2}_{GW}\, \rho_{rad}\, \Big{(}
\frac{a}{a_{kin}} \Big{)}^2, \label{GWd} \ee where with the use of
(\ref{keyGW}) we finally obtain the following constraint from
gravitational waves \be \frac{\rho_{g}}{\rho_{rad}}|_{a=a_{eq}}=\frac{64}{3\pi}\,
h^{2}_{GW}\, \frac{2\, H_{kin}^{2/3\alpha_{1}}\,
\Gamma^{1/3\alpha_{2}}}{m^{2-1/\alpha_{1}}\, \sigma_{i}^{2}\,
\beta_{1}^{5/6\alpha_{1}}\,\beta_{2}^{1/6\alpha_{2}}} \ll 1~. \label{GWconstr2} \ee
For the case $1b.$ this constraint becomes \be
\frac{\rho_{g}}{\rho_{rad}}|_{a=a_{eq}}=\frac{64}{3\pi}\, h^{2}_{GW}\, \frac{2\,
H_{kin}^{2/3\alpha_{1}}\,
\Gamma^{1/3\alpha_{1}}}{m^{2-1/\alpha_{1}}\, \sigma_{i}^{2}\,
\beta_{1}^{1/\alpha_{1}}} \ll 1~, \label{GWconstr1} \ee
while for case $3b.$ it is \be
\frac{\rho_{g}}{\rho_{rad}}|_{a=a_{eq}}=\frac{64}{3\pi}\, h^{2}_{GW}\, \frac{2\,
H_{kin}^{2/3\alpha_{2}}\,
\Gamma^{1/3\alpha_{2}}}{m^{2-1/\alpha_{2}}\, \sigma_{i}^{2}\,
\beta_{2}^{1/\alpha_{2}}} \ll 1~. \label{GWconstr3} \ee
The constraint on decay parameter $\Gamma$ is now given by \be
H_{eq} < \Gamma < m~. \ee

\section{Curvaton Dynamics in Randall-Sundrum Brane-world Model}

We will apply the general formalism developed in Sect.~2 in the
Randall-Sundrum Brane-world model \cite{RScosmology}. In this
model we have two regimes before nucleosynthesis: a
five-dimensional regime and a four-dimensional one.

\subsection{The Randall-Sundrum Brane-world Model}

The gravitational action of the Randall-Sundrum Brane-world model
is \be \label{RS2Action} S = \frac{1}{2\kappa^{2}_{5}}\, \int
d^{5}x \, \sqrt{-^{(5)}g}\,\, \Big{(}
^{(5)}\mathcal{R}\,-2\Lambda_{5}\,\, \Big{)}  +
\frac{1}{2\kappa^{2}_{4}}\, \int_{y=0} d^{4}x \, \sqrt{-^{(4)}g}\,
\, \Big{(}-2\Lambda_{4}\Big{)}~, \ee where
$\kappa_{5}^{2}=M_{5}^{-3}=8\pi G_{5}$ is the $5D$ fundamental
gravitational constant while $\kappa_{4}^{2}=M_{4}^{-2}=8\pi
G_{4}$ is the effective $4D$ gravitational constant. $\Lambda_{5}$
is the cosmological constant of the AdS bulk which is related to
its characteristic lengthscale $l$ by $\Lambda_{5} = - 6/l^2$ and
$\Lambda_{4}$ is the $4D$ cosmological constant, while the brane
tension is $\lambda=\Lambda_{4} / \kappa^2_{4}$. The brane tension
is related  to the AdS lengthscale through the fine-tuning
relation \be \label{RSTuning}
\lambda=\frac{6}{\kappa_{5}^{2}\,l}~, \ee which implies that
$l=\kappa_{5}^2/\kappa_{4}^2$ indicating that $l$ plays the role
of the crossover scale between the $5D$ and the $4D$ regimes. The
Friedmann equation is given by  \be \label{RS2Fried} H^2 =
\frac{\kappa_{4}^{2}}{3}\, \rho\, \Big{(} 1+\frac{\rho}{2\lambda}
\Big{)}~, \ee which gives rise to two different regimes for the
dynamical evolution of the brane universe
\begin{itemize}
\item
 the RS regime, when $\rho
\gg \lambda\,\,\, (Hl \gg 1)\,\, $ \be \label{RS2Regime}
H^2\simeq\frac{\kappa_{5}^4}{36}\, \rho^2, \ee
\item
the 4D regime, when $ \rho \ll \lambda\,\,\, (Hl \ll 1)\,\,\, $
 \be \label{GRRegime}
H^2\simeq\frac{\kappa_{4}^2}{3}\, \rho~. \ee
\end{itemize}

\subsection{Slow-roll Inflation in the Randall-Sundrum Brane-world Model}

We will now review the results of the slow-roll inflationary
dynamics for the Randall-Sundrum brane-world model
\cite{MaartensChaoticRS2}. We consider an inflaton field $\phi$
with energy density given by (\ref{inflaton}) and obeying the
Klein-Gordon equation (\ref{inflmotion}). Then
 the slow-roll conditions for
the inflaton field are \bea \label{slowroll}
\frac{1}{2} \dot{\phi}^{2} &\ll& V(\phi)~, \nonumber \\
\ddot{\phi} &\ll& 3\, H \, \dot{\phi}~, \eea and the slow-roll
parameters, at high energies, are defined as \bea
\label{slowrollparamepsilonRS2} \varepsilon &=&
\frac{1}{2\kappa_{4}^{2}}\, \Big{(} \frac{V'}{V} \Big{)} ^{2}
\, \Big{[} \frac{4\lambda}{V} \Big{]}~, \\
\label{slowrollparametaRS2} \eta &=& \frac{1}{\kappa_{4}^{2}}\,
\Big{(} \frac{V''}{V} \Big{)}\, \Big{[} \frac{2\lambda}{V}
\Big{]}~, \eea where terms in brackets express the $5D$
corrections to general relativity. The number of e-folds is given
by \be N = - \kappa_{4}^{2} \int^{\phi_{f}}_{\phi_{i}}
\frac{V}{V'}\, \Big{[} \frac{V}{2\lambda} \Big{]}\,  d \phi~. \ee
The scalar amplitude and the scalar spectral index generated
during inflation have been computed \cite{MaartensChaoticRS2} (see
also~\cite{Langlois:2005nd}) and are respectively given by \bea
A_{s}^{2} &=&
\frac{\kappa_{4}^{6}}{75\pi^{2}}\,\frac{V^{3}}{V'^{2}}\,
\Big{[}G_{RS}^{2}(Hl)\Big{]}
\Big{|}_{\tilde{k}=aH}~, \\
n_{s} -1 &=& \frac{d\, \ln{A_{s}^2}}{d\, \ln{\tilde{k}}} = 2 \eta
- 6 \varepsilon~, \eea where  \be G^{2}_{RS}(x)=\Big{[}
\frac{\sqrt{1+x^2}+1}{2} \Big{]} ^{3}, \ee is the RS correction
and $\tilde{k}$ is the comoving wavenumber. We  find at the high
energy limit $(\rho \gg \lambda)$  \be A_{s}^{2} =
\frac{\kappa_{4}^{6}}{75\pi^{2}}\,\frac{V^{3}}{V'^{2}}\, \Big{[} 1
+ \frac{V}{2\lambda} \Big{]}^{3}.\label{scapert} \ee Moreover the
tensor amplitude and tensor spectral index are respectively given
by \cite{MaartensATRS2} \bea A_{T}^{2} &=& \frac{32\,
\kappa_{4}^{4}}{75\pi^{2}}\, V\,
\Big{[} F_{RS}^2(Hl) \Big{]}\,\Big{|}_{\tilde{k}=aH}~,\label{tenampl} \\
n_{T} &\equiv& \frac{d\, \ln{A_{T}^2}}{d\, \ln{\tilde{k}}} \simeq
-2\varepsilon~, \eea with \be F^{2}_{RS}(x)= \Big{\{}
\sqrt{1+x^2}-x^2\, \ln \Big{[} \frac{1}{x} +
\sqrt{1+\frac{1}{x^2}} \Big{]} \Big{\}} ^{-1/2}. \ee The tensor
amplitude (\ref{tenampl}) in the high energy limit becomes \be
A_{T}^{2} = \frac{32\, \kappa_{4}^{4}}{75\pi^{2}}\, V\, \Big{[}
\frac{V}{2\lambda} \Big{]}\,\Big{|}_{\tilde{k}=aH}~, \ee giving
the consistency relation  \be \frac{A^{2}_{T}}{A_{s}^{2}}\simeq
-\frac{n_{T}}{2}~. \ee

For a chaotic inflation with inflaton potential
$V(\phi)=\frac{1}{2}\, m_{\phi}^2\, \phi^2$, $N=55$ and assuming
that the density perturbations are generated by the inflaton field
using $A_{s}\simeq 2 \times 10^{-5}$ the above relations give \bea
\label{ChaoticRS2}
m_{\phi} &\simeq& \frac{0.32}{(192\,N+80)^{5/6}}\, M_{5} \simeq 1.4\,\times 10^{-4}\, M_{5}~, \\
\lambda &\simeq& 5.6\,\times 10^3\, (192N+80)^5 \, \frac{m^6}{M_{4}^{2}} \simeq 4 \,\times 10^{-38} \, M_{5}^{6}~, \\
\phi_{i} &\simeq& \frac{(192N+80)^{2/3}}{0.57}\, M_{5} \simeq 8.5\,\times 10^2\, M_{5}~,\label{phi} \\
\phi_{f} &\simeq& 3.8\, \times 10^2\, M_{5}~,\label{phf} \\
ns &\simeq& 1-\frac{5}{2N} \simeq 0.9545~.\label{specindex} \eea
All the above parameters of the RS model except the scalar
spectral index are functions of the five-dimensional fundamental
mass. To keep the quantum gravity corrections under control we
require  $M_5 < 10^{17}$ GeV. Also note that the value of the
scalar spectral index is within the limits of the three-year WMAP
data.

\subsection{Curvaton Reheating in the Randall-Sundrum Brane-world Model}

Inflation occurs in the five-dimensional RS regime and depending
on which regime the curvaton reheating occurs we have six
different cases for the curvaton evolution: \bea
Case\, 1a&:& \hspace{0.5in} H_{RS}>H_{f}>m>H_{eq1}>\Gamma>H_{RS.GR}>H_{nucl} \nn \\
Case\, 1b&:& \hspace{0.5in} H_{RS}>H_{f}>m>\Gamma>H_{eq1}>H_{RS.GR}>H_{nucl} \nn \\
Case\, 2a&:& \hspace{0.5in} H_{RS}>H_{f}>m>H_{RS.GR}>H_{eq2}>\Gamma>H_{nucl} \nn \\
Case\, 2b&:& \hspace{0.5in} H_{RS}>H_{f}>m>H_{RS.GR}>\Gamma>H_{eq2}>H_{nucl} \nn \\
Case\, 3a&:& \hspace{0.5in} H_{RS}>H_{f}>H_{RS.GR}>m>H_{eq2}>\Gamma>H_{nucl} \nn \\
Case\, 3b&:& \hspace{0.5in}
H_{RS}>H_{f}>H_{RS.GR}>m>\Gamma>H_{eq2}>H_{nucl}~. \eea
If we apply the general results of section $2$ we obtain: \\
\be \label{RSCurvaton54} H_{RS.GR}= \frac{\lambda\,
\kappa_{5}^{2}}{3}~, \ee and for the various cases we have
\begin{itemize}
\item
Case 1a.~: \,\,\,\,\,
  \be \label{RSCurvaton1a}  H_{eq1} = \frac{\sigma_{i}^4}{4}\,
\frac{m^3}{36\, M_{5}^{6}}~, \ee
\item
Case 2a.~: \,\,\,\,
   \be
\label{RSCurvaton2a}  H_{eq2} = \frac{1}{12} \frac{m\,
\sigma_{i}^2}{M_{5}^{3}}\, \sqrt{\frac{\lambda\, m}{3\, M_{5}^{3}}}~, \ee
\item
Case 3a.~: \,\,\,\,\
 \be \label{RSCurvaton3a}  H_{eq2} =
\frac{\kappa_{4}^2}{6}\, \sigma_{i}^2\, m~, \ee

\item
Case 1b.~: \,\,\,\,\,
  \be \label{RSCurvaton1b}  H_{eq1} = \frac{\sigma_{i}^6}{8}\,
\frac{m^{9/2}}{216\, M_{5}^{9}\,\Gamma^{1/2}}~, \ee
\item
Case 2b.~: \,\,\,\,
   \be
\label{RSCurvaton2b}  H_{eq2} = \frac{3^{3/4}}{216} \frac{m^2\,
\sigma_{i}^3\,\lambda}{M_{5}^{6}}\, \Big{(}\frac{m}{\lambda \, M_{5}^{3}}\Big{)}^{1/4}\,\Gamma^{-1/2}~, \ee
\item
Cases 3b.~: \,\,\,\,\
 \be \label{RSCurvaton3b}  H_{eq2} =
\frac{\sigma_{i}^3\, m^{3/2}}{6^{3/2}\, M_4^3}\,\Gamma^{-1/2}~. \ee
\end{itemize}

As for the gravitational waves, using (\ref{GWeq}) we have the
following constraint for the Cases 1a. and 2a. \be
\frac{\rho_{g}}{\rho_{\sigma}}|_{a=a_{eq}}=\frac{32}{3\pi}\, h_{GW}^2\, \Big{(}
\frac{6\, H_{kin}} {m\, \sigma_{i}^{2}\, \kappa_{4}^{2}} \Big{)}
^{2/3} \ll 1~,
 \label{RSGWeq12}
\ee and for Case 3a. the constraint \be
\frac{\rho_{g}}{\rho_{\sigma}}|_{a=a_{eq}}=\frac{32}{3\pi}\, \frac{h_{GW}^2}{m}\,
\Big{(} \frac{12\, H_{kin}^{1/2}} {\sigma_{i}^{2}\,
\kappa_{5}^{2}} \Big{)} ^{2/3} \ll 1~. \label{RSGWeq3}
\ee Similarly, using  (\ref{GWconstr1}), (\ref{GWconstr2}) and
(\ref{GWconstr3}) we obtain respectively for cases $1b.$, $2b.$
and $3b.$ the constraints \bea
\frac{\rho_{g}}{\rho_{rad}}|_{a=a_{eq}}&=&\frac{128}{\pi}\, h^{2}_{GW}\,
\frac{\Gamma^{1/3}\, H_{kin}^{2/3}}{m\, \sigma_{i}^{2}\,
\kappa_{4}^2} \ll 1 \label{RSGWconstr1}~, \\
\frac{\rho_{g}}{\rho_{rad}}|_{a=a_{eq}}&=&\frac{128\times 2^{5/6}}{\pi}\, h^{2}_{GW}\,
\frac{\Gamma^{1/3}\, H_{kin}^{1/3}}{m^{3/2}\, \sigma_{i}^{2}\,
\kappa_5^{5/3}\,\kappa_{4}^{1/3}} \ll 1 \label{RSGWconstr2}~, \\
\frac{\rho_{g}}{\rho_{rad}}|_{a=a_{eq}}&=&\frac{512}{\pi}\, h^{2}_{GW}\,
\frac{\Gamma^{1/6}\, H_{kin}^{1/3}}{m^{3/2}\, \sigma_{i}^{2}\,
\kappa_{5}^2} \ll 1 \label{RSGWconstr3}~. \eea

\subsection{Constraints on the Parameters in the Randall-Sundrum Brane-world Model}

In all cases the transition from five to four dimensions using
(\ref{RSCurvaton54}) happens at \be H_{RS.GR}=7.5\times 10^{-39}
M^{3}_{5}~.  \ee Also from the nucleosynthesis constraint
$\Gamma>H_{nucl}=10^{-40}M_{Pl}$ we obtain \be \Gamma> 1.2\times
10^{-21}~GeV~, \ee for $M_{Pl}=1.2\times 10^{19}~GeV$, while
inflation ends at \be H_{f}=\frac{\kappa_{5}^{2}}{12} m^{2}_{\phi}
\phi^{2}_{f}=2.4\times 10^{-4}~M_{5}~. \ee

We will analyse in more details the case 3 which is more
interesting. The other two cases give similar bounds for the
parameters. For the  case 3 the universe inflates in 5D and the
primordial density perturbations are generated by the inflaton
field. Looking at the values of the inflaton field (\ref{phi}) and
(\ref{phf}) at the beginning and at the end of inflation we see
that during inflation the inflaton field does not change much.
Therefore we can assume that it survives without decay when the
universe enters the 4D regime in which the curvaton field reheats
the universe. In this picture, inflation and the generation of
density perturbations are high energy effects, while reheating can
occur at lower energy scales.

We find for both Cases $3a.$ and $3b.$ the following constraints
on the  parameters
 \bea 5.98\times
10^{5}~GeV~\,\,\,\, <~ M_{5}~ &<& 10^{17}~GeV \nn \\
1.6\times10^{-21}~GeV\,\,\,\, <~ \Gamma\, <\,m~ &<& 7.5\times 10^{12}~GeV \nn \\
5.73\times10^{2}~GeV\,\,\,\,\, <~ \sigma_{i}~ &<& 3.92\times 10^{19}~GeV \nn \\
2.46\times10^{18}~GeV^{3}\,\,\,\,\, <~ m \sigma^{2}_{i}~ &<&
1.15\times 10^{52}~GeV^{3}. \label{results}
\eea
The lower bounds of the mass of curvaton field and its initial value indicate that
the curvaton field could be identified with a low energy scalar
particle.

If the density perturbations are generated by the inflaton field
then relations (\ref{RSGWeq3}) and (\ref{RSGWconstr3}) give
respectively for cases $3a.$ and $3b.$ the following extra
constraints on the different parameters \bea
m\,\sigma_i^{4/3} &>& 1.80\times10^{-44}\,M_5^{13/3} \\
m^{3/2}\,\sigma_i^{2} &>&
2.85\times10^{-43}\,M_5^{16/3}\,\Gamma^{1/6}. \eea One obtains
similar constraints on the parameters $m, \sigma_i, M_5$ and
$\Gamma$ for cases $1$ and $2$.

 If the density perturbations are not generated by the inflaton
 then, the mass, the initial and final values of the inflaton
 fields as a function of $A_{s}$ and $M_{5}$ are
 \bea
 m_{\phi}&=&9.44 \times 10^{-2} A_{s}^{2/3}M_{5}~, \nn \\
 \phi_{i}&=&3.31\times 10~ A_{s}^{-1/3}M_{5}~, \nn \\
 \phi_{f}&=&1.02\times 10~ A_{s}^{-1/3}M_{5}~. \eea
If the amplitude of density perturbations $A_{s}$ generated by the
inflaton field
 is much lower than its observed value and for fixed values of
the inflaton field (in order to keep the successful prediction of
the spectral index (\ref{specindex})) the fundamental mass $M_{5}$
could get values much lower than the four-dimensional Planck
mass~\footnote{This opens up the possibility that inflation could
occur in a low energy regime~\cite{dimoLyth}.}. However if the
curvaton is to generate the density perturbations additional
constrains should be satisfied by the curvaton parameters
\cite{curvaton}. In the case that curvaton decays after
domination, the following constraint should be satisfied
\begin{equation}
{\cal P}_\zeta \simeq \frac{1}{9\pi^2} \frac{H^2_{{\rm
i}}}{\sigma^2_{{\rm i}}} \,~, \label{3e24}
\end{equation}
where ${\cal P}_\zeta$ is the Bardeen parameter~\footnote{The
Bardeen parameter is used to describe the curvature perturbations.
If the perturbations are adiabatic, which is usually the case for
a single scalar field, then the curvature perturbations are
identified with the scalar perturbations through the formula
$A^{2}_{s}=4{\cal P}_\zeta/25$.} whose observed value is about $2
\times 10^{-9}$ and $H_i$ is the value of the Hubble parameter at
curvaton oscillation. Hence for case $3a.$ we find that \be
m=3\sqrt{6}\,\pi\, M_{Pl}\,P_{\zeta}^{1/2}~, \ee which gives
$m=1.24\times 10^{16} GeV$ for $P_{\zeta}=2\times 10^{-9}$ which
is incompatible with (\ref{results}). Thus for case $3a$ density
perturbations could not have been generated by the curvaton field.
On the contrary, for cases $1a.$ and $2a$ density perturbations
can be generated by the curvaton field, because the extra
constrainted relation involving the parameters $m, \sigma_{i}$ and
$M_5$  is compatible with the corresponding bounds of the
parameters for the cases $1a.$ and $2a.$. If the curvaton decays
when subdominant, the Bardeen parameter is given by
\cite{curvaton}
\begin{equation}
{\cal P}_\zeta \simeq \frac{r^2_{{\rm d}}}{36\pi^2}
\frac{H^2_{{\rm i}}}{\sigma^2_{{\rm i}}} \,~. \label{3e25}
\end{equation}
Here $r_{{\rm d}}$ is the ratio of curvaton energy density to
stiff scalar matter at curvaton decay, which is given for case
$3b.$ by~\cite{liddle}
\begin{equation}
r_{{\rm d}} = \left. \frac{\rho_\sigma}{\rho_\phi}
\right|_{a=a_{{\rm d}}} \!\!\! = \frac{m\,\sigma^2_{{\rm
i}}}{6\,\Gamma\,M^2_{\rm Pl}} \, ~. \label{3e26}
\end{equation}
This gives the following constraint \be m\, \sigma_i = 9.22 \times
10^{26}\, \Gamma^{1/2}\, GeV^2, \ee which is compatible with the
 results of (\ref{results}). For cases $1b.$ and $2b.$ we
respectively have \bea r_d &=&
\frac{m^{3/2}\,\sigma_i^2}{12\,\Gamma^{1/2}\,M_5^3}~, \\
r_d &=& \frac{m^{3/2}\,\sigma_i^2}{12\,\Gamma\,M_5^3}\,
H_{RS.GR}^{1/2}~,\eea which in each case gives an extra constraint
that is compatible with the various bounds for cases $1b.$ and
$2b$. Therefore in all cases $b$, the scalar perturbations can be
generated by the curvaton field.

\section{Curvaton Dynamics in Gauss-Bonnet Brane-world Model}

\subsection{The Gauss-Bonnet Brane-world Model}

In this section we will study the curvaton dynamics in the
Gauss-Bonnet brane-world
model which has the
following gravitational action \bea \label{GBAction} S &=&
\frac{1}{2\kappa^{2}_{5}}\, \int d^{5}x \, \sqrt{-^{(5)}g}\,\,
\Big{(}  ^{(5)}\mathcal{R}\,-2\Lambda_{5}\,\, \nonumber \\
&+&\alpha\, \Big{(}^{(5)}\mathcal{R}^2-4\,
^{(5)}\mathcal{R}_{ab}\, ^{(5)}\mathcal{R}^{ab} +
^{(5)}\mathcal{R}_{abcd}\,
 ^{(5)}\mathcal{R}^{abcd}\Big{)}\Big{)} \nonumber \\
&+& \frac{1}{2\kappa^{2}_{4}}\, \int_{y=0} d^{4}x \,
\sqrt{-^{(4)}g}\, \, \Big{(}-2\Lambda_{4}\Big{)}~, \eea where
$\alpha
> 0$ is the Gauss-Bonnet coupling constant. At first order in
$\alpha$, the relation between the $5D$ cosmological constant of
the AdS bulk and its characteristic lengthscale $l$ has the form
\be \label{GBL5} \Lambda_{5} = - \frac{6}{l^2} + \frac{12
\alpha}{l^4} = - \frac{6}{l^2}\, \Big{(}1-\frac{1}{2}\beta
\Big{)}~, \ee where $\beta \equiv 4\alpha/l^2$. The brane tension
$\lambda=\Lambda_{4} / \kappa^2_{4}$ is related to the AdS
lengthscale through the
 fine-tuning relation \be \label{GBTuning}
\lambda=\frac{6}{\kappa_{5}^{2}\,l}\, \Big{(}1-\frac{1}{3}\beta
\Big{)}~, \ee implying that $l=\kappa_{5}^2/\kappa_{4}^2\,
(1+\beta)$ which can be considered as a crossover scale between
the 5D and the 4D regimes.

 The Friedmann equation of the Gauss-Bonnet brane model is given by
  \be \label{GBFried}
\kappa_{5}^{2}\, (\rho + \lambda) = \frac{2}{l}\,
\sqrt{1+H^{2}\,l^{2}}\, \Big{(}3+\beta
\,(2H^{2}\,l^{2}-1)\Big{)}~, \ee which, assuming that the GB term
represents correction to the Einstein-Hilbert term, i.e. $\beta
\ll 1$, gives rise to three different regimes for the dynamical
evolution of the brane-universe. We use a characteristic
Gauss-Bonnet energy scale
\be \label{GBscale}
m_{\beta} = \Big{[} \frac{8(1-\beta)^3}{l^2 \beta \kappa_{5}^4} \Big{]}^{1/8},
\ee
and we find that:
\begin{itemize}
\item
at the GB high energy regime, $\rho \gg m_{\beta}^{4}\,\,\,(Hl
\gg \beta^{-1} \gg 1)$ \be \label{GBRegime} H^2\simeq \Big{[}
\frac{\kappa_{5}^2}{4\beta l^2}\, \rho \Big{]} ^{2/3}, \ee
\item at an
intermediate RS regime, $ m_{\beta}^{4} \gg \rho \gg
\lambda\,\,\, (\beta^{-1} \gg Hl \gg 1)$ \be \label{RSGBRegime}
H^2\simeq\frac{\kappa_{4}^2}{6\lambda}\, \rho^{2}~, \ee \item at
the GB low energy regime (GR limit), $\rho \ll \lambda\,\,\,(\beta^{-1} \gg 1 \gg
Hl)$\be \label{GRGBRegime}
H^2\simeq\frac{\kappa_{4}^2}{3}\, \rho~. \ee
\end{itemize}

\subsection{Slow-roll Inflation in the Gauss-Bonnet Brane-world Model}

We will now review the results of the slow-roll inflationary
dynamics for the Gauss-Bonnet brane-world model
\cite{MaartensChaoticGB}. If we define a new variable $\chi$ as
\be \label{GBFriedFinal} \kappa_{5}^{2}(\rho +
\lambda)=\frac{2}{l} \Big{[} \frac{2(1-\beta)^{3}}{\beta} \Big{]}
^{1/2} \, \sinh\, \chi~, \ee then the Friedmann equation
(\ref{GBFried}) can be written in the particularly simple form
\cite{LidseyInflationGB} \be \label{LidseyFriedGB}
H^2=\frac{1}{\beta l^2}\, \Big{[}(1-\beta)\, \cosh\Big{(}
\frac{2\chi}{3} \Big{)} -1 \Big{]}~. \ee As previously we consider
an inflaton field $\phi$ with energy density given by
(\ref{inflaton}) and obeying the Klein-Gordon equation
(\ref{inflmotion}). Since during slow-roll inflation $V \approx
\rho \gg \lambda$ then from (\ref{GBscale}) and
(\ref{GBFriedFinal}) we have that \be \label{GBPotential}
V=m_{\beta}^{4}\, \sinh\, \chi~. \ee The slow-roll parameters, at
high energies, for the inflaton field satisfying the slow-roll
conditions (\ref{slowroll}) are \bea
\label{slowrollparamepsilonGB} \varepsilon_{GB} &=&
\varepsilon_{RS}\, \frac{2(1-\beta)^{4}\, \sinh \frac{2}{3} \chi\,
\tanh \chi\, \sinh^{2} \chi}
{9(1+\beta)(3-\beta)[(1-\beta)\, \cosh \frac{2}{3} \chi -1]^{2}}~, \\
\label{slowrollparametaGB} \eta_{GB} &=& \eta_{RS}
\frac{2(1-\beta)^{3}\, \sinh^{2} \chi}
{3(1+\beta)(3-\beta)[(1-\beta)\, \cosh \frac{2}{3} \chi -1]}~,
\eea where $\varepsilon_{RS}$ and $\eta_{RS}$ are the RS slow-roll
parameters (\ref{slowrollparamepsilonRS2}) and
(\ref{slowrollparametaRS2}). In the limit $\chi \ll 1$ we have
$\varepsilon_{GB} \rightarrow \varepsilon_{RS}$ and $\eta_{GB}
\rightarrow \eta_{RS}$ recovering the RS results. The number of
e-folds are \be \label{efoldsGB} N \approx -3
\int^{\chi_{f}}_{\chi_{i}} \frac{H^2}{dV/d\chi}\, \Big{(}
\frac{d\phi}{d\chi}\Big{)}^{2}\, d\chi~, \ee where $\chi_{f}$ is
evaluated at the end of inflation $(\varepsilon = 1)$  and
$\chi_{i}$ is evaluated when cosmological scales leave the
horizon. This latter parameter is constrained by a quantum gravity
upper limit which requires that $V < (8\pi\,M_{5})^4$ thus
obtaining \be \label{GBqgconstraint} \sinh\, \chi_{i} < \Big{(}
\frac{8\pi\,M_{5}}{m_{\beta}} \Big{)} ^4. \ee

The scalar amplitude and the scalar
spectral index generated during inflation have been computed
\cite{MaartensScalarGB} and are respectively
given by \bea A_{s}^{2} &=&
\frac{\kappa_{4}^{6}}{75\pi^{2}}\,\frac{V^{3}}{V'^{2}}\, \Big{[}
G_{GB}^{2}(Hl)\Big{]}
\Big{|}_{\tilde{k}=aH}~, \\
n_{s} -1 &=& \frac{d\, \ln{A_{s}^2}}{d\, \ln{\tilde{k}}} = 2 \eta
- 6 \varepsilon~, \eea where \be G^{2}_{GB}(x)=\Big{[}
\frac{3(1+\beta)\,x^2}{2\sqrt{1+x^2}\,(3-\beta+2\beta
x^2)+2(\beta-3)} \Big{]} ^{3} \ee is the GB correction and
$\tilde{k}$ is the comoving wavenumber. We then find at the GB
regime \be \label{scalarGB} A_{s}^{2} =
\frac{\kappa_{4}^{6}}{75\pi^{2}}\,\frac{V^{3}}{V'^{2}}\, \Big{[}
\frac{27}{64}\, \Big{(} \frac{1+\beta}{\beta} \Big{)} ^{3}\,
\frac{1}{(Hl)^3} \Big{]}~. \ee Moreover the tensor amplitude and
tensor spectral index are respectively given by
\cite{MaartensScalarGB} \bea A_{T}^{2} &=& \frac{32\,
\kappa_{4}^{4}}{75\pi^{2}}\, V\,
\Big{[} F_{GB}^2(Hl) \Big{]}\,\Big{|}_{\tilde{k}=aH}~, \\
n_{T} &\equiv& \frac{d\, \ln{A_{T}^2}}{d\, \ln{\tilde{k}}} \simeq
-2\varepsilon \, \Big{[} 1-\frac{(H\,l)^2\, F_{GB}^2(H\,l)\,
\{1-(1-\beta)\, \sqrt{1+x^2}\,\sinh^{-1}\,(H\,l)^{-1}
\}}{(1+\beta)\, \sqrt{1+x^2}} \Big{]},\,\, \eea with \be
F^{-2}_{GB}(x)= \sqrt{1+x^2}-\Big{(} \frac{1-\beta}{1+\beta}
\Big{)} \,x^2\, \sinh^{-1}\, \frac{1}{x}~. \ee We then find that
at the GB regime \be A_{T}^{2} = \frac{32\,
\kappa_{4}^{4}}{75\pi^{2}}\, V\, \Big{[}
\frac{1+\beta}{2\beta}\Big{]}\, (Hl)^{-1}, \ee and the consistency
relation becomes  \be \frac{A^{2}_{T}}{A_{s}^{2}}\simeq
-\frac{Q_{GB}}{2}\, n_{T}~, \ee where \be
Q_{GB}=\frac{1+\beta+2\beta\,x^2}{1+\beta+\beta\, x^2}~. \ee
Inflation can end either in the GB regime $(\chi_{f} \gg 1)$ or in
the RS regime $(\chi_{f} \ll 1)$. We consider first the case of a
chaotic inflation with potential $V(\phi)=\frac{1}{2}\,
m_{\phi}^2\, \phi^2$ \textbf{\textit{ending in the RS regime}}. We
have assumed that $\beta \ll 1$ and from the definitions of
paragraph $4.1.$ we can write the following relations \bea
\label{RedefGB}
\kappa_{5}^4 &=& \frac{6\, \kappa_{4}^2}{\lambda}~, \\
m_{\beta}^8 &=& \frac{2\, \lambda^2}{9\, \beta}~, \\
l^2\, \kappa_{5}^4 &=& \frac{36}{\lambda^2}~. \eea The slow-roll
parameters (\ref{slowrollparamepsilonGB}) and
(\ref{slowrollparametaGB}) are then \bea
\label{slowrollparamepsilonGB1}
\varepsilon &=& \frac{4 \beta \, m_{\phi}^2}{3 \kappa_{4}^2\, \lambda}\, f(\chi)~, \\
\eta &=& \frac{2 \beta \, m_{\phi}^2}{\kappa_{4}^2\, \lambda}\,
g(\chi)~, \label{slowrollparametaGB1} \eea where \bea f(\chi) &=&
\frac{\sinh(\frac{2\chi}{3})\,\tanh \chi}
{\Big{[}\cosh(\frac{2\chi}{3})-1\Big{]}^2}~, \\
g(\chi) &=& \frac{1}{\cosh(\frac{2\chi}{3})-1}~. \eea The number
of e-folds (\ref{efoldsGB}) is \be \label{efoldsGB1} N =
\frac{\kappa_{4}^{2}\, \lambda}{4\beta\, m_{\phi}^2}\,
\Big{[}I(\chi_{i})-I(\chi_{f})\Big{]}~, \ee with \be I(\chi) =
\frac{3}{2} \Big{[}\cosh(\frac{2\chi}{3})-1
-\ln\Big{(}\frac{1}{3}+\frac{2}{3}\,\cosh(\frac{2\chi}{3})\Big{)}
\Big{]}~, \ee whereas the amplitude of scalar perturbations
(\ref{scalarGB}) is given by \be \label{scalarGB1} A_s^2 =
\frac{\kappa_{4}^6\, \lambda^2}{32\, \sqrt{2}\, \pi^2\,
\beta^{5/2}\, m_{\phi}^2}\,
\frac{\Big{[}\cosh(\frac{2\chi}{3})-1\Big{]}^3}{\sinh\, \chi}~.
\ee From (\ref{slowrollparamepsilonGB1}) the condition for the end
of inflation $\varepsilon_{f} = 1$ together with (\ref{efoldsGB1})
and with $\chi_f \ll 1$ gives the following relation \be
\label{relationGB1} \frac{6\beta\,m_{\phi}^2}{\kappa_{4}^2\,
\lambda} = \frac{I(\chi_{i})}{\Big{(}1+\frac{2}{3}\,N\Big{)}}~,
\ee while (\ref{scalarGB1}) gives \be \label{relationGB2}
\frac{\beta^{3/2}}{\kappa_{4}^{4}\, \lambda} =
\frac{3\,\Big{(}1+\frac{2}{3}\,N\Big{)}} {16\,\sqrt{2}\, \pi^2\,
A_{s}^2\, I(\chi_{i})}\,\,
\frac{\Big{[}\cosh(\frac{2\chi}{3})-1\Big{]}^3}{\sinh\, \chi}~.
\ee Moreover using (\ref{relationGB1}) with the slow-roll
parameters (\ref{slowrollparamepsilonGB1}) and
(\ref{slowrollparametaGB1}) we get the following expression for
the scalar spectral index \be \label{spectralGB} n_s =
1-\frac{2}{3}\,\frac{I(\chi_i)}{1+\frac{2}{3}\,N}\,[2\,f(\chi_i) -
g(\chi_i)] \ee and the condition (\ref{GBqgconstraint}) becomes
\be \label{GBqgconstraint1} \sinh^6\, \chi <
\frac{9\,\beta^3}{128\, \kappa_4^8\, \lambda^2}~. \ee We thus have
three equations (\ref{relationGB1}), (\ref{relationGB2}) and
(\ref{spectralGB}) with $6$ parameters, $n_s, N, \chi_i, \beta,
\lambda$ and $m_{\phi}$. Moreover using \be \lambda = 6
\frac{M_5^6}{M_{4}^2}~, \ee we get the following final expressions
for $\beta, m_{\phi}, m_{\beta}$ and $\phi_i$ as functions of $N,
\chi_i$ and $M_5$ \bea \label{GBbetafinal} \beta &=& \Big{[}
\frac{9\Big{(}1+\frac{2}{3}\,N\Big{)}}
{8\,\sqrt{2}\,\pi^2\,A_{s}^2\,I(\chi_{i})}\Big{]}^{2/3}\,\,
\frac{\Big{[}\cosh\Big{(}\frac{2\chi_i}{3}\Big{)}-1\Big{]}^2}{\sinh\chi_i^{2/3}}\,\frac{M_5^4}{M_4^4}~, \\
m_{\phi}&=& \Big{[} \frac{8\,\sqrt{2}\,\pi^2\,A_{s}^2\,I(\chi_{i})^{5/2}}
{9\,\Big{(}1+\frac{2}{3}\,N\Big{)}^{5/2}}\Big{]}^{1/3}\,\,
\frac{\sinh\chi_i^{1/3}}{\cosh\Big{(}\frac{2\chi_i}{3}\Big{)}-1}\,M_5~, \label{GBmphifinal}\\
m_{\beta}&=& \Big{[} \frac{256\,\pi^2\,A_{s}^2\,I(\chi_{i})}
{9\,\Big{(}1+\frac{2}{3}\,N\Big{)}}\Big{]}^{1/12}\,\,
\frac{\sinh\chi_i^{1/12}}{\Big{[}\cosh\Big{(}\frac{2\chi_i}{3}\Big{)}-1\Big{]}^{1/4}}\,M_5~, \label{GBmbetafinal}\\
\phi_i&=& \Big{[} \frac{144\Big{(}1+\frac{2}{3}\,N\Big{)}^4}
{\pi^2\,A_{s}^2\,I(\chi_{i})^4}\Big{]}^{1/6}\,\,
\Big{[}\cosh\Big{(}\frac{2\chi_i}{3}\Big{)}-1\Big{]}^{1/2}\,\,\sinh\chi_i^{1/3}\,M_5~.
\label{GBphifinal} \eea The constraint relation
(\ref{GBqgconstraint1}) with $M_4=M_{Pl}=1.2\times 10^{19}$ GeV,
 $A_s^2 = 4\times\,10^{-10}$ and $N=55$  gives
$\chi_{i} < 12.57$.

\begin{figure}[h!]
\centering
\includegraphics*[scale=1]{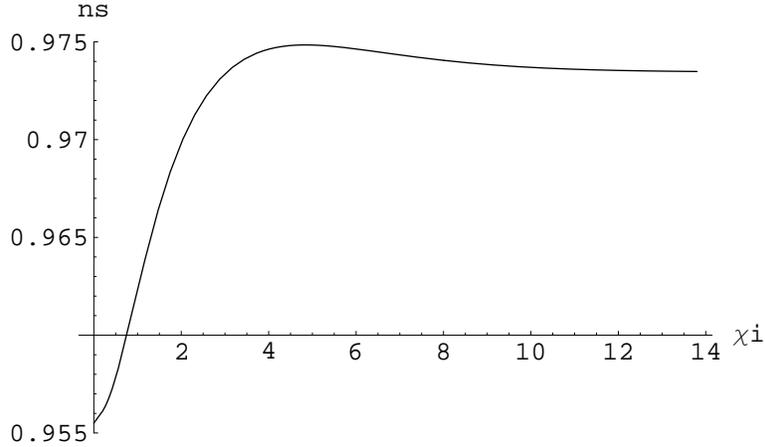}
\caption{Scalar spectral index as a function of $\chi_i$ for $N=55$.}
\end{figure}

In Fig. $1$ we plot the scalar spectral index $n_s$ as a function
of $\chi_{i}$. Motivated  from the WMAP three-year results
\cite{3jwmap} we take $n_s < 0.966$ therefore we get $\chi_i <
1.42$. As a consequence inflation starts either at a RS regime for
$\chi_i \ll 1$ (where we recover the results of section $3$) or
during a mixed GB-RS regime for $\chi_i \sim 1$. We can thus fix
$\chi_i = 0.8$ giving $n_s = 0.9604$. Moreover we have in our case
$\phi_{i} \simeq 3.34 \phi_f$. Therefore we obtain the following
values for the different parameters as functions of $M_5$ \bea
\beta &=& 2.31\times10^{-71}\,M_5^4~, \\
m_{\beta}&=& 2.53\times10^{-1}\,M_5~, \\
m_{\phi}&=& 6.63\times10^{-5}\,M_5~, \\
\lambda&=&4.16\times10^{-38}\,M_5^6~, \\
\phi_i&=& 1.28\times10^3\,M_5~, \\
\phi_f&=& 3.83\times10^2\,M_5~. \eea The parameter $\beta \ll 1$,
therefore for various values of $\beta$ we get

\vspace{0.3in}
\begin{tabular}{|c|c|c|c|}
  \hline
  $\beta=10^{-3}$ & $\lambda^{1/4}=1.04\times10^{16}$ GeV & $m_{\phi}=5.40\times10^{12}$ GeV
  & $m_\beta=2.05\times10^{16}$ GeV\\
  \hline
  \,\, & $\phi_i=1.04\times10^{20}$ GeV & $\phi_f=3.12\times10^{19}$ GeV & $M_5 = 8.11\times10^{16}$ GeV \\
  \hline
  $\beta=10^{-5}$ & $\lambda^{1/4}=1.86\times10^{15}$ GeV & $m_{\phi}=1.71\times10^{12}$ GeV
  & $m_\beta=6.49\times10^{15}$ GeV\\
  \hline
  \,\, & $\phi_i=3.29\times10^{19}$ GeV & $\phi_f=9.85\times10^{18}$ GeV & $M_5 = 2.57\times10^{16}$ GeV \\
  \hline
  $\beta=10^{-7}$ & $\lambda^{1/4}=3.30\times10^{14}$ GeV & $m_{\phi}=5.40\times10^{11}$ GeV
  & $m_\beta=2.05\times10^{15}$ GeV\\
  \hline
  \,\, & $\phi_i=1.04\times10^{19}$ GeV & $\phi_f=3.12\times10^{18}$ GeV & $M_5 = 8.11\times10^{15}$ GeV \\
  \hline
  $\beta=10^{-12}$ & $\lambda^{1/4}=4.40\times10^{12}$ GeV & $m_{\phi}=3.03\times10^{10}$ GeV
  & $m_\beta=1.15\times10^{14}$ GeV\\
  \hline
  \,\, & $\phi_i=5.84\times10^{17}$ GeV & $\phi_f=1.75\times10^{17}$ GeV & $M_5 = 4.56\times10^{14}$ GeV \\
  \hline
\end{tabular}
\vspace{0.3in}

We can see from the above table that in order to keep the value of
the inflaton field below the Planck scale, the $\beta$ parameter
should be $ < 10^{-7}$ constraining in this way the value of the
fundamental mass to  $M_5 < 10^{16}$ GeV.

We  consider also the case of a chaotic inflation with potential
$V(\phi)=\frac{1}{2}\, m_{\phi}^2\, \phi^2$ \textbf{\textit{ending
in the GB regime}} where $\chi_f \gg 1$. From
(\ref{slowrollparamepsilonGB1}) the condition for the end of
inflation $\varepsilon_{f} = 1$ together with (\ref{efoldsGB1})
gives the following relation \be \label{relationGB21}
\frac{6\beta\,m_{\phi}^2}{\kappa_{4}^2\, \lambda} =
\frac{3\,I(\chi_{i})}{2\,\Big{(}N+\frac{1}{2}\Big{)}}~, \ee while
(\ref{scalarGB1}) gives \be \label{relationGB211}
\frac{\beta^{3/2}}{\kappa_{4}^{4}\, \lambda} =
\frac{2\,\Big{(}N+\frac{1}{2}\Big{)}} {16\,\sqrt{2}\, \pi^2\,
A_{s}^2\, I(\chi_{i})}\,\,
\frac{\Big{[}\cosh(\frac{2\chi}{3})-1\Big{]}^3}{\sinh\, \chi}~.
\ee Moreover using (\ref{relationGB21}) with the slow-roll
parameters (\ref{slowrollparamepsilonGB1}) and
(\ref{slowrollparametaGB1}) we get the following expression for
the scalar spectral index \be \label{spectralGB11} n_s =
1-\frac{I(\chi_i)}{N+\frac{1}{2}}\,[2\,f(\chi_i) - g(\chi_i)] \ee
and the condition (\ref{GBqgconstraint}) becomes \be
\label{GBqgconstraint111} \sinh^6\, \chi < \frac{9\,\beta^3}{128\,
\kappa_4^8\, \lambda^2}~. \ee We thus have three equations
(\ref{relationGB21}), (\ref{relationGB211}) and
(\ref{spectralGB11}) with $6$ parameters, $n_s, N, \chi_i, \beta,
\lambda$ and $m_{\phi}$. Taking again $N=55$ the constraint
relation (\ref{GBqgconstraint111}) gives $\chi_{i} < 12.56$. The
plot of the scalar spectral index $n_s$ as a function of
$\chi_{i}$ has the same shape as in Fig. $1$. In this case however
we have $\chi_f \gg 1$ which means that the value of the spectral
index should be greater than the largest value given by the
 WMAP three-year results \cite{3jwmap}. Therefore the case
of an inflationary period ending in a GB regime is ruled out by
observational data. Inflation should end in a RS
regime~\footnote{An alert reader would have noticed that if we
relax the upper bound of the spectral index obtained from the
three-year WMAP results and consider larger values of
 $n_{s}$,~ Fig. 1 shows that $\chi_{f}$ could be much larger than 1, and then
we also have the interesting case of inflation to end in the GB
regime.}.

\subsection{Curvaton Reheating in the Gauss-Bonnet Brane-world Model}

For the Gauss-Bonnet brane-world model we have three different
regimes of cosmological evolution and the use of the curvaton
reheating mechanism gives the following cases: \be Case
1:\hspace{0.5in} H_{GB}>H_{GB.RS}>H_{f}~, \ee where $H_{f}$ is the
Hubble parameter at the end of inflation. We consider that the GB
regime dominates at very early time and is followed by the
intermediate RS regime where inflation takes place. Thus inflation
ends at a RS regime and we recover the case of inflation results
and curvaton reheating in the RS model (section $3$). \be Case\,
2:\hspace{0.5in} H_{GB}>H_{f}>H_{GB.RS}~. \ee In this case
inflation occurs during the GB regime which is immediately
followed by a transition to the intermediate RS regime. Thus in
this case inflationary parameters are determined by the results of
subsection $4.2$ while the curvaton reheating mechanism gives the
same results as in subsection $3.3$.

 We can also consider that
inflation and oscillation of the curvaton occur at the GB regime,
thus we can use the results of subsection $4.2$ for the
inflationary parameters and we have three different cases for the
curvaton evolution: \bea
Case\, 3a&:& \hspace{0.5in} H_{GB}>H_{f}>m>H_{eq1}>\Gamma>H_{GB.RS}>H_{RS.GR}>H_{nucl} \nn \\
Case\, 3b&:& \hspace{0.5in} H_{GB}>H_{f}>m>\Gamma>H_{eq1}>H_{GB.RS}>H_{RS.GR}>H_{nucl} \nn \\
Case\, 4a&:& \hspace{0.5in} H_{GB}>H_{f}>m>H_{GB.RS}>H_{eq2}>\Gamma>H_{RS.GR}>H_{nucl} \nn \\
Case\, 4b&:& \hspace{0.5in} H_{GB}>H_{f}>m>H_{GB.RS}>\Gamma>H_{eq2}>H_{RS.GR}>H_{nucl} \nn \\
Case\, 5a&:& \hspace{0.5in} H_{GB}>H_{f}>m>H_{GB.RS}>H_{RS.GR}>H_{eq3}>\Gamma>H_{nucl} \nn \\
Case\, 5b&:& \hspace{0.5in}
H_{GB}>H_{f}>m>H_{GB.RS}>H_{RS.GR}>\Gamma>H_{eq3}>H_{nucl}~. \eea

If we apply the general results of section $2$ we obtain: \\
\bea \label{GBCurvaton45} H_{GB.RS}&=&
\frac{\sqrt{6}}{2\sqrt{\beta}\, l}
~, \\
H_{RS.GR}&=& \frac{2}{l(1+\beta)} \label{GBCurvaton54}
~. \eea Then for the various cases we have:
\begin{itemize}
\item
Case 3a.~: \be \label{GBCurvaton3a} H_{eq1} =
\Big{(}\frac{\kappa_{5}^{2}}{8\beta\,l^{2}}\Big{)}^{2/3}
\sigma_{i}^{4/3}\, m^{1/3}
. \ee
\item
Case 4a.~: \be \label{GBCurvaton4a} H_{eq2} =
\frac{\sigma_{i}^4}{64}\, \frac{\kappa_{5}^{4}}{\beta\, l^{2}}\, m
~. \ee
\item
Case 5a.~ In these case we perform similar calculations
as in subsection $2.1..2$ but with two transitions before the
equivalence, and we find \be \label{GBCurvaton5a} H_{eq3} =
\frac{\sqrt{3}\, \sigma_{i}^{2}\, m^{1/2}\,\kappa_5 \kappa_{4}}{12\,\beta^{1/2}\, l}~. \ee
\item
Case 3b.~: \be \label{GBCurvaton3b} H_{eq1} =
\frac{\kappa_{5}^{2}}{8\beta\,l^{2}} \sigma_{i}^{2}\, m^{1/2}\, \Gamma^{-1/2}
. \ee
\item
Case 4b.~: \be \label{GBCurvaton4b} H_{eq2} =
\frac{\sigma_{i}^6}{64\sqrt{6}}\, \frac{\kappa_{5}^{6}\,m^{3/2}}{\beta^{3/2}\, l^{3}}\, \Gamma^{-1/2}
~. \ee
\item
Case 5b.~ In these case we perform similar calculations
as in subsection $2.2..2$ but with two transitions before the
equivalence, and we find \be \label{GBCurvaton5b} H_{eq3} =
\frac{\sigma_{i}^{3}\, m^{3/4}}{24 \beta^{3/4}\, l^{3/2}}\, \kappa_{Pl}^{3/2}\, \kappa_{5}^{3/2}\, \Gamma^{-1/2}~. \ee
\end{itemize}

As for the gravitational waves we have the following constrained
relations: For cases 3a., 4a. and 5a., the kination and
oscillation epochs are in the GB regime, thus using (\ref{GWeq})
we obtain \be \frac{\rho_{g}}{\rho_{\sigma}}|_{a=a_{eq}}=\frac{128}{3\pi}\, h_{GW}^2\,
H_{kin}\, \Big{(} \frac{\beta\, l^2}{m^{1/2}\, \sigma_{i}^{2}\,
\kappa_{5}^{2}} \Big{)} ^{2/3} \ll 1~. \label{GBGWeq}
\ee For cases 3b., 4b. and 5b. we need the regimes when kination
and oscillation take place and in which regime the curvaton starts
decaying. Using equation (\ref{GWconstr2}) we obtain: \bea Case\,
3b&:& \hspace{0.5in} \frac{\rho_{g}}{\rho_{rad}}|_{a=a_{eq}}=\frac{512}{3\pi}\,
h^{2}_{GW}\, H_{kin}\, \frac{\beta\, l^2}{m^{1/2}\,
\sigma_{i}^{2}\, \kappa_{5}^{2}}
\, \Gamma^{1/2}  \ll 1~, \label{GBGWconstr3} \\
Case\, 4b&:& \hspace{0.5in}
\frac{\rho_{g}}{\rho_{rad}}|_{a=a_{eq}}=\frac{128}{3\pi}\times 6144^{1/6}\, h^{2}_{GW}\, H_{kin}\,
\frac{\beta^{5/6}\, l^{5/3}}{m^{1/2}\, \sigma_{i}^{2}\, \kappa_{5}^{2}}
\, \Gamma^{1/6}  \ll 1~, \label{GBGWconstr4} \\
Case\, 5b&:& \hspace{0.5in}
\frac{\rho_{g}}{\rho_{rad}}|_{a=a_{eq}}=\frac{128}{3\pi}\times 3072^{1/6}\, h^{2}_{GW}\, H_{kin}\,
\frac{\beta^{5/6}\, l^{5/3}}{m^{1/2}\, \sigma_{i}^{2}\, \kappa_{5}^{2}}
\, l^{1/6}\, \Gamma^{1/3}  \ll 1~. \label{GBGWconstr5} \eea

\subsection{Constraints on the Parameters in the Gauss-Bonnet Brane-world Model}

In all cases the transition from GB to  RS and from RS to GR using
(\ref{GBCurvaton45}) and (\ref{GBCurvaton54})  happens at \bea
H_{GB.RS}&=& \frac{8.51}{\sqrt{\beta}}\times10^{-39}\,M_5^3~, \\
H_{RS.GR}&=& 1.39\times10^{-38}\,M_5^3~.  \eea Also from the
nucleosynthesis constraint $\Gamma>H_{nucl}=10^{-40}M_{Pl}$ we
obtain \be \Gamma> 1.2\times 10^{-21}~GeV~, \ee for
$M_{Pl}=1.2\times 10^{19}~GeV$. From the results obtained in
section $4.2$  cases $2, 3, 4$ and $5$  are ruled out and the only
valid cases are $1a.$ and $1b.$ studied in section $3$ where
inflation ends in a RS regime with \be
H_{f}=\frac{\kappa_{5}^{2}}{12} m^{2}_{\phi}
\phi^{2}_{f}\simeq5.37\times 10^{-5}\,M_5~, \ee for $\chi_i =
0.8$.

 If the density perturbations are not generated
by the inflaton
 then, the mass, the initial and final values of the inflaton
 fields as a function of $A_{s}$ and $M_{5}$ are
 \bea
 m_{\phi}&=&9.00 \times 10^{-2} A_{s}^{2/3}M_{5}~, \nn \\
 \phi_{i}&=&3.47\times 10~ A_{s}^{-1/3}M_{5}~, \nn \\
 \phi_{f}&=&1.04\times 10~ A_{s}^{-1/3}M_{5}~, \eea
and for the gravitational waves we get similar constraints as in
section $3.4$.

As in the RS case, if the density perturbations are generated by
the fluctuations of the curvaton field then we get low energy
values for the fundamental mass $M_{5}$. It can be also shown that
the constraints (\ref{3e24}) and (\ref{3e25}) give similar results
as in section $3.4$.

\section{Curvaton Dynamics in Induced Gravity (DGP) Brane-world Model}

\subsection{The Induced Gravity Brane-world Model}

In this section we will apply the curvaton reheating mechanism to
the Induced Gravity (DGP) brane-world
scenario which has
the following gravitational action \bea \label{INDAction}
S &=& \frac{1}{2\kappa^{2}_{5}}\, \int d^{5}x \, \sqrt{-^{(5)}g}\,\,
 \Big{(} ^{(5)}\mathcal{R}-2\Lambda_{5}\,\, \Big{)}
\nn \\
&+& \frac{r}{2\kappa^{2}_{5}}\, \int_{y=0} d^{4}x \,
\sqrt{-^{(4)}g}\, \, \Big{(}^{(4)}\mathcal{R}\,-2\Lambda_{4}\,\,
\Big{)}~, \eea where $r = \kappa_{5}^{2}/\kappa_{4}^{2} > 0$ is
the induced gravity crossover scale, one of the two characteristic
lengthscales of the model, while the second characteristic
lenghtscale, the AdS lengthscale, is  the one related to the
Planck coupling constant as $l = \kappa_{5}^{2}/\kappa_{Pl}^{2}$.
As in the Randall Sundrum model, the relation between the $5D$
cosmological constant of the AdS bulk and its characteristic
lengthscale $l$ is \be \label{INDL5} \Lambda_{5} = -
\frac{6}{l^2}~. \ee The brane tension $\lambda=\Lambda_{4} /
\kappa^2_{4}$ is related to the AdS lengthscale through the
fine-tuning relation \be \label{INDTuning}
\lambda=\frac{6}{\kappa_{5}^{2}\,l}~. \ee Moreover, in order to
recover GR at low energies the two $4D$ coupling constants are
related by $\kappa_{Pl}^{2}=\mu\, \kappa_{4}^{2}$, where $\mu =
\kappa_{4}^{2}\, r^{2}\, \lambda/6$~\cite{lpapzamchaotic}. Two
different cosmological evolutions can be distinguished. The first,
for $\mu \gg 1$, is a pure $4D$ evolution at all energies which we
will not discuss it here. The second, for $\mu \ll 1$, is giving
an interesting cosmological evolution that we will consider. The
Friedmann equation is given by  \be \label{INDFried}
H^{2}=\frac{\kappa_{4}^{2}}{3}\, \Big{\{} 1 +
\frac{6}{\kappa_{4}^{2}\, r^{2}\, \rho} + \frac{\lambda}{\rho} +
\epsilon\, \frac{2\sqrt{3}}{\kappa_{4}\, r\, \rho^{1/2}}\, \Big{(}
1 + \frac{3}{\kappa_{4}^{2}\, r^{2}\, \rho} + \frac{\lambda}{\rho}
+ \frac{\lambda}{\rho}\, \frac{\kappa_{4}^{2}\, r^{2}\,
\lambda}{12} \Big{)} ^{1/2} \Big{\}}~, \ee where $\epsilon = \pm
1$. However it was shown in \cite{lpapzamchaotic} that only for
$\epsilon = -1$ we get an inflationary phase. Using the Friedmann
equation (\ref{INDFried}) we can distinguish  three different
regimes for the dynamical evolution of the brane universe:
\begin{itemize}
\item
the IND regime, when $Hl \gg Hr \gg 1$ \be \label{INDRegime}
H^2\simeq \frac{\kappa_{4}^{2}\, \rho}{3} \Big{(} 1 -
\frac{2\sqrt{3}}{\kappa_{4}\, r\, \rho^{1/2}} \Big{)}~, \ee
\item
the intermediate RS regime, when $Hl \gg 1 \gg Hr$ \be
\label{RSINDRegime} H^2\simeq \frac{\kappa_{Pl}^2}{6\lambda}\,
\rho^{2}~, \ee
\item
the GR regime at low energies, when $Hr \ll Hl \ll 1$ \be
\label{GRINDRegime} H^2\simeq\frac{\kappa_{Pl}^2}{3}\, \rho~. \ee
\end{itemize}
Notice that in the case of $r \longrightarrow 0$ we recover the RS
model, whereas if we also have $l \longrightarrow 0$ we recover GR
cosmology.

\subsection{Slow-roll Inflation in the Induced Gravity Brane-world Model}

We will review the results of the slow-roll inflationary dynamics
for the Induced Gravity brane-world model \cite{lpapzamchaotic}.
As before we consider an inflaton field $\phi$ with energy density
given by (\ref{inflaton}) and obeying the Klein-Gordon equation
(\ref{inflmotion}).  The slow-roll conditions for the inflaton
field are \bea \label{slowrollIND}
\dot{\phi}^{2}\Big{(}1-\frac{\sqrt{3}/2}{\kappa_{4} r\,V^{1/2}}
\Big{)}&<&V\Big{(}1+\epsilon\frac{2\sqrt{3}}{\kappa_{4}
r\,V^{1/2}} \Big{)}~, \\
\ddot{\phi} &\ll& 3\, H \, \dot{\phi}~, \eea and the slow-roll
parameters, at high energies, are found to be \bea
\label{slowrollparamepsilonIND} \varepsilon &=& \varepsilon_{GR}\,
\Big{(}1+\frac{3\sqrt{3}} {\kappa_{4}
r\,V^{1/2}}\Big{)}~, \\
\label{slowrollparametaIND} \eta &=& \eta_{GR} \Big{(}
1+\frac{2\sqrt{3}}{\kappa_{4} r\,V^{1/2}} \Big{)}~, \eea where
$\varepsilon_{GR}$ and $\eta_{GR}$ are the GR slow-roll
parameters. The number of e-folds are \be
N=-\frac{\kappa^{2}_{Pl}}{\mu}\int^{\phi_{f}}_{\phi_{i}}\frac{V}{V^{\prime}}
\Big{(}1-\frac{2\sqrt{3}}{\kappa_{4} r\,V^{1/2}}\Big{)}
d\phi~,\label{efolgrIND}\ee where $\phi_{f}$ is evaluated at the
end of inflation $(\varepsilon = 1)$, and $\phi_{i}$ is evaluated
when cosmological scales leaves the horizon. The scalar amplitude
and the scalar spectral index generated during inflation have been
computed \cite{MaartensPertIND} and are respectively given by \bea
A_{s}^{2} &=&
\frac{\kappa_{4}^{6}}{75\pi^{2}}\,\frac{V^{3}}{V'^{2}}\, \Big{[}
G_{IND}^{2}(Hl)\Big{]}
\Big{|}_{\tilde{k}=aH}~, \\
n_{s} -1 &=& \frac{d\, \ln{A_{s}^2}}{d\, \ln{\tilde{k}}} = 2 \eta
- 6 \varepsilon~, \eea where \be
G^{2}_{IND}(x)=\frac{x^6}{\Big{\{}
\mu\,x^2-2(1-\mu)[1-\sqrt{1+x^2}]\Big{\}}^3}~, \label{indexIND} \ee is the IND
correction and $\tilde{k}$ is the comoving wavenumber. We then
find at the IND regime \be A_{s}^{2} =
\frac{\kappa_{4}^{6}}{75\pi^{2}}\,\frac{V^{3}}{V'^{2}}\, \Big{[}
\frac{1}{\mu^3} \Big{]}~. \ee Moreover the tensor amplitude and
tensor spectral index are respectively given by
\cite{MaartensPertIND} \bea A_{T}^{2} &=&
\frac{32\,\kappa_{4}^{4}}{75\pi^{2}}\, V\,
\Big{[} F_{IND}^2(Hl) \Big{]}\,\Big{|}_{\tilde{k}=aH}~, \\
n_{T} &\equiv& \frac{d\, \ln{A_{T}^2}}{d\, \ln{\tilde{k}}}~, \eea
with \be F^{2}_{IND}(x)= \mu+(1-\mu)\Big{[} \sqrt{1+x^2}-x^2\,
arcsinh\, \frac{1}{x}\Big{]}~. \ee We then find that at the IND
regime \be A_{T}^{2} = \frac{32\,\kappa_{4}^{4}}{75\pi^{2}}\, V\,
\Big{[} \frac{1}{\mu}\Big{]}~, \ee so that the consistency
relation becomes at IND regime \be
\frac{A^{2}_{T}}{A_{s}^{2}}\simeq -Q_{IND}\frac{n_{T}}{2}~, \ee
where $Q_{IND}$ is a complicated differential equation between $V$
and $F_{IND}$~.

For a chaotic inflation with potential $V(\phi)=\frac{1}{2}\,
m_{\phi}^2\, \phi^2$ the following expressions  can be obtained
using relations (\ref{slowrollparamepsilonIND})-(\ref{indexIND})
 \bea \phi_{i}&=&\frac{15+7\,c}{2\,(1-n_{s})\,
\kappa_{4}^2}\Big{(}1+
\sqrt{1-\frac{112\,N\,(1-n_{s})}{(1-c)\,(15+7\,c)^2}}\Big{)}~, \\
V_0&=&\frac{m_{\phi}^2}{2}=\frac{3\,\pi^2\,(1-c)}{1.25\times10^7\,\kappa_4^4\,\phi_i^2\,
\Big{(}12N-\kappa_4^2\,\phi_i^2\,(1-c)\,(1+3\,c)\Big{)}}~,\\
r&=&\frac{M_4^2}{M_5^3}=-\frac{4\sqrt{3}\,\kappa\,\phi_i\,(1-c)}
{\sqrt{V_0}\Big{(}4N-\kappa_4^2\,\phi_i^2\,(1-c^2)\Big{)}}~, \eea
which are essentially functions of the spectral index $n_s$, the parameter
$c=\phi_f / \phi_i$, the four-dimensional coupling constant $\kappa_{4}$ and the number
of e-folds $N$. Fixing the value of the scalar spectral index to
$n_s=0.9645$, the values of the various parameters using the above
relations for $N=55$, $c=0.1$ and $\mu=0.001$ are \bea \label{ChaoticIND}
m_{\phi} &\simeq& 2.53\times10^{12} GeV~, \\
M_{5} &\simeq& 4.07\times10^{15} GeV~, \\
\lambda^{1/4} &\simeq& 1.17\times10^{14}, \\
\phi_{i} &\simeq& 5.97\times10^{18} GeV~, \label{INDphi}\\
\phi_{f} &\simeq& 0.1\, \phi_{i} = 5.97\times10^{17} GeV~. \label{INDphf} \eea

\subsection{Curvaton Reheating in the Induced Gravity Brane-world Model}

In the Induced Gravity Model we have three different cosmological
regimes. We distinguish the following cases for the curvaton
reheating mechanism\footnote{Similar analysis is carried out
in~\cite{Zhang:2006np}.}: \be Case\, 1:\hspace{0.5in}
H_{IND}>H_{IND.RS}>H_{f}~, \ee where $H_{f}$ is the Hubble
parameter at the end of inflation. We consider that the IND regime
dominates at very high energies and is followed by the
intermediate RS regime where inflation takes place. Thus inflation
ends at a RS regime and we recover the case of inflation results
and curvaton reheating in the RS model (section $3$). \be Case\,
2:\hspace{0.5in} H_{IND}>H_{f}>H_{IND.RS}~, \ee where we consider
that inflation occurs during the IND regime which is immediately
followed by a transition to the intermediate RS regime. Thus in
this case inflationary parameters are determined by the results of
subsection $5.2$ while for the curvaton reheating mechanism the
results of  subsection $3.3$ are valid.

 If inflation and
oscillation of the curvaton occurs at the IND regime, results of
subsection $5.2$ are considered for the inflationary parameters
and we have three different cases for the curvaton evolution: \bea
Case\, 3a&:& \hspace{0.5in} H_{IND}>H_{f}>m>H_{eq1}>\Gamma>H_{IND.RS}>H_{RS.GR}>H_{nucl} \nn \\
Case\, 3b&:& \hspace{0.5in} H_{IND}>H_{f}>m>\Gamma>H_{eq1}>H_{IND.RS}>H_{RS.GR}>H_{nucl} \nn \\
Case\, 4a&:& \hspace{0.5in} H_{IND}>H_{f}>m>H_{IND.RS}>H_{eq2}>\Gamma>H_{RS.GR}>H_{nucl} \nn \\
Case\, 4b&:& \hspace{0.5in} H_{IND}>H_{f}>m>H_{IND.RS}>\Gamma>H_{eq2}>H_{RS.GR}>H_{nucl} \nn \\
Case\, 5a&:& \hspace{0.5in} H_{IND}>H_{f}>m>H_{IND.RS}>H_{RS.GR}>H_{eq3}>\Gamma>H_{nucl} \nn \\
Case\, 5b&:& \hspace{0.5in}
H_{IND}>H_{f}>m>H_{IND.RS}>H_{RS.GR}>\Gamma>H_{eq3}>H_{nucl}~.
\eea

If we apply the general results of section $2$ we obtain: \bea
\label{INDCurvaton45}
H_{IND.RS}&=& 2\,r~, \\
H_{RS.GR}&=& 2\, l=2\,\mu\,r \label{INDCurvaton54}\eea and for
cases $3a.$ and $3b.$ respectively \bea \label{INDCurvaton3a}
H_{eq1} &=&
\frac{\kappa_{Pl}^{2}}{6\mu}\, \sigma_{i}^{4/3}\, m^{1/3}~, \\
\label{INDCurvaton3b} H_{eq1} &=&
\frac{\kappa_{Pl}^{3}}{(6\mu)^{3/2}}\, \sigma_{i}^{3}\, m^{3/2}\, \Gamma^{-1/2}~,\eea
while for cases $4a.$ and $4b.$ we obtain \bea
\label{INDCurvaton4a} H_{eq2} &=& \frac{\kappa_{Pl}^4}{72\mu^2}\, \,
\sigma_{i}^4\,m^2\,r~, \\
\label{INDCurvaton4b} H_{eq2} &=& \frac{\kappa_{Pl}^6}{432\sqrt{2}\,\mu^3}\, \,
\sigma_{i}^6\,m^3\,r^{3/2}\,\Gamma^{-1/2}~.\eea For the cases $5a.$ and $5b.$ we
perform similar analysis as in subsection $2.2$ but with two
transitions before the equivalence, and we find respectively
\bea
\label{INDCurvaton5a} H_{eq3} &=&
\frac{\kappa_{Pl}^{2}}{6\mu}\, \sigma_{i}^{2}\,m~, \\
\label{INDCurvaton5b} H_{eq3} &=&
\frac{\kappa_{Pl}^{3}}{6\sqrt{6}\mu^{3/4}}\, \sigma_{i}^{3}\,m^{3/2}\,\Gamma^{-1/2}~. \eea

For the gravitational waves the most interesting cases are
 cases $3a.$, $4a.$ and $5a.$ in which the  kination and oscillation epochs are in the IND regime, thus
using (\ref{GWeq}) we obtain \be
\frac{\rho_{g}}{\rho_{\sigma}}|_{a=a_{eq}}=\frac{32}{3\pi}\, h_{GW}^2\, H_{kin}^{2/3}\,
\frac{6^{2/3}\,\mu^{1/6}}{m^{2/3}\, \sigma_{i}^{1/3}\, \kappa_{Pl}^{1/3}}~ \ll 1.
\label{INDGWeq} \ee For cases $3b.$, $4b.$ and $5b.$ we need the
regimes when kination and oscillation take place and in which
regimes the curvaton starts decaying. Using equation
(\ref{GWconstr2}) we obtain: \bea Case\, 3b&:& \hspace{0.5in}
\frac{\rho_{g}}{\rho_{rad}}|_{a=a_{eq}}=\frac{128}{\pi}\, h^{2}_{GW}\,
H_{kin}^{2/3}\, \frac{\mu}{m\, \sigma_{i}^{2}\, \kappa_{Pl}^{2}}
\, \Gamma^{1/3}  \ll 1~, \label{INDGWconstr3} \\
Case\, 4b&:& \hspace{0.5in} \frac{\rho_{g}}{\rho_{rad}}|_{a=a_{eq}}=\frac{128}{\pi}\,
h^{2}_{GW}\, H_{kin}^{2/3}\, \frac{\mu}{m\, \sigma_{i}^{2}\,
\kappa_{Pl}^{2}}
\, \Big{(} \frac{2\,\Gamma}{r}\Big{)}^{1/6}  \ll 1~, \label{INDGWconstr4} \\
Case\, 5b&:& \hspace{0.5in} \frac{\rho_{g}}{\rho_{rad}}|_{a=a_{eq}}=\frac{256}{\pi}\,
h^{2}_{GW}\, H_{kin}^{1/3}\, \frac{\mu}{m^{3/2}\, \sigma_{i}^{2}\,
\kappa_{Pl}^{2}\, r^{5/6}} \, \Gamma^{1/3}  \ll 1~.
\label{INDGWconstr5} \eea

\subsection{Constraints on the Parameters in the Induced Gravity Brane-world Model}

In all cases the transition from four to five dimensions and from five to four
dimensions using respectively (\ref{INDCurvaton45}) and (\ref{INDCurvaton54})
happens at
\bea
H_{IND.RS}&=&2.88\times10^{38}\,M_5^{-3}\simeq4.27\times10^{-12}\,\, GeV~, \\
H_{RS.GR}&=&2.88\times10^{38}\,\mu\,M_5^{-3}\simeq4.27\times10^{-15}\,\,
GeV~, \eea where the $M_5$ was determined for $n_s=09645,~ N=55,~
c=0.1$ and $\mu=0.001$. Also from the nucleosynthesis constraint
$\Gamma>H_{nucl}=10^{-40}M_{Pl}$ we obtain \be \Gamma> 1.2\times
10^{-21}~GeV~, \ee for $M_{Pl}=1.2\times 10^{19}~GeV$. Moreover
inflation ends at \be H_{f}=\frac{\kappa_{4}}{\sqrt{6}}
m_{\phi}\,\phi_{f}\,\sqrt{\Big{(}1-
\frac{2\sqrt{6}}{\kappa_4\,r\,m_{\phi}\,\phi_f}\Big{)}}=5.09\times
10^{13}\,\,GeV~, \ee thus $H_f > H_{IND.RS}$ which invalidates
case $1$ and because $H_f \gg H_{IND.RS}$  the transition to RS
regime cannot occur immediately after the end of inflation,
therefore case $2$ is also excluded. Moreover, the low energy and
close to each other values of $H_{IND.RS}$ and $H_{RS.GR}$ make
case $3$ to be the most probable case for the reheating mechanism.
In case $3$
 the universe inflates in 4D induced gravity regime and the
primordial density perturbations are generated by the inflaton
field. Also the curvaton starts to oscilate in the
four-dimensional induced gravity high energy regime. To have a
better fitting of the parameters we fix $c=0.1$ from which the
values of the inflaton field (\ref{INDphi}) and (\ref{INDphf}) at
the begining and at the end of inflation do not change much.

Using (\ref{INDCurvaton3a})-(\ref{INDCurvaton5b}) together with
(\ref{ChaoticIND})-(\ref{INDphf}) we find the following
constraints on the parameters for
\begin{itemize}
\item Case $3a.$ \bea
4.27\times10^{-12}~GeV\,<\, \Gamma \,&<& \, m \, < \,5.09\times10^{13}~GeV \nn \\
3.68\times10^{24}~GeV^{5/3}\, <\, 8.62\times10^{35}\,\Gamma \,
&<&\, \sigma_{i}^{4/3}\,m^{1/3}\, < \,4.39\times
10^{49}~GeV^{5/3}, \label{results3a}\eea \item Case $3b.$ \bea
4.27\times10^{-12}~GeV\,<\, \Gamma \,< \, m \, &<& \,5.09\times10^{13}~GeV \nn \\
3.58\times10^{-6}< 4.06\times10^{11}\,\Gamma^{3/2} <\, \sigma_{i}\,m^{1/2}\,
&<&\, 1.50\times10^{16}\,\Gamma^{1/6} \, < \,2.89\times 10^{18}~, \label{results3b}\eea
\item Case $4a.$ \bea
4.27\times10^{-15}~GeV\,<\, \Gamma \,< \, 4.27\times10^{-12}\,&<&\,m \, < \,5.09\times10^{13}~GeV \nn \\
2.99\times10^{69}~GeV^{6}\, <\, 2.21\times10^{84}\,\Gamma \, <\,
\sigma_{i}^{4}\,m^{2}\, &<& \,2.99\times 10^{72}~GeV^{6},
\label{results4a}\eea \item Case $4b.$ \be Incompatible
\label{results4b}\ee \item Case $5a.$ \bea
4.27\times10^{-12}~GeV\,<\, m \, &<& \,5.09\times10^{13}~GeV \nn \\
1.20\times10^{-21}~GeV\,<\, \Gamma \,&<& \, 4.27\times10^{-15}~GeV \nn \\
1.04\times10^{15}~GeV^{3}\, <\, 8.64\times10^{35}\,\Gamma \, &<&\,
\sigma_{i}^{2}\,m\, < \,3.70\times 10^{21}~GeV^{3},
\label{results5a}\eea \item Case $5b.$ \be Incompatible
\label{results5b}\ee
\end{itemize}

Moreover, for cases $3a.$, $4a.$ and $5a.$ the gravitational
constraint (\ref{INDGWeq}) gives $m^2\,\sigma_i > 8.00 \times
10^{27} \, GeV^{3}$. For case $3b.$ relation (\ref{INDGWconstr3})
gives $m\,\sigma_i^2 > 1.93 \times 10^{39} \, \Gamma^{1/3}$ while
cases $4b.$ and $5b.$ are excluded from (\ref{results4b}) and
(\ref{results5b}). If the density perturbations are not generated
by the inflaton then, the different parameters expressed as a
function of $A_{s}$ are \bea
m_{\phi}&=&1.29 \times 10^{17} A_{s}~, \nn \\
M_5 &=&1.50\times 10^{17}~ A_{s}^{1/3}~, \nn \\
\lambda^{1/4}&=&2.62\times10^{16}~ A_{s}^{1/2}~. \eea As in the
case of the RS or the GB model, if density perturbations are
generated by the fluctuations of the curvaton field, the
fundamental mass $M_{5}$ can get low energy values. However, the
constraints (\ref{3e24}) and (\ref{3e25}) should also be satisfied
for the curvaton parameters of the induced gravity model. A
detailed analysis shows, that only the case $3b.$ can satisfy
these constraints at the expense of introducing two extra
constraints on the parameters $m, \sigma_i$ and $\Gamma$ \bea
m\,\sigma_i &>& 8.70
\times 10^{29}\,\, GeV^2, \\
m\,\sigma_i &>& 4.84 \times 10^{25}\, \Gamma^{1/2}\, GeV^2. \eea
We note that case $5b.$ is excluded in both cases, whether density
perturbations are generated by the inflaton field or by the curvaton field.

\section{Curvaton Dynamics in Gauss-Bonnet and Induced Gravity (GBIG) Brane-world Model}

\subsection{The GBIG Brane-world Model}

In this section we will apply the curvaton reheating mechanism to
the Gauss-Bonnet and Induced Gravity (GBIG) brane-world
\cite{papa} scenario which has the following gravitational action
\bea \label{INDAction}
S = \frac{1}{2\kappa^{2}_{5}}\, \int d^{5}x \, \sqrt{-^{(5)}g}\,\, \Big{(}  ^{(5)}\mathcal{R}\,-2\Lambda_{5}\,\, \\
+\alpha\, (^{(5)}\mathcal{R}^2-4\, ^{(5)}\mathcal{R}_{ab}\,
^{(5)}\mathcal{R}^{ab}
 + ^{(5)}\mathcal{R}_{abcd}\, ^{(5)}\mathcal{R}^{abcd})\Big{)} \\
+ \frac{r}{2\kappa^{2}_{5}}\, \int_{y=0} d^{4}x \,
\sqrt{-^{(4)}g}\, \, \Big{(}
^{(4)}\mathcal{R}-2\Lambda_{4}\Big{)}~, \eea where $\alpha>0$ is
the Gauss-Bonnet coupling constant, $r =
\kappa_{5}^{2}/\kappa_{4}^{2} > 0$ the induced gravity crossover
scale, where $\kappa_{4}$ is the effective $4D$ coupling constant
 different from the $4D$ Planck coupling constant
$\kappa_{Pl}$. The AdS lengthscale is given by $l =
\kappa_{5}^{2}/\kappa_{Pl}^{2}$. Here the relation between the
$5D$ cosmological constant of the AdS bulk and its characteristic
lengthscale $l$ is as in the GB model \be \label{GBIGL5}
\Lambda_{5} = - \frac{6}{l^2}+\frac{12\alpha}{l^4}~, \ee with the
constraint $\alpha \leq l^2/4$ which gives $\Lambda_5 < 0$. The
general form of the Friedmann equation as a cubic equation in
$H^2$ is given by \be \label{GBIGCubic} 4\Big{[} 1 +\frac{8}{3}\,
\alpha\, \Big{(} H^2 + \frac{\Phi}{2} \Big{)}
\Big{]}^2\,(H^2-\Phi)
=\Big{[}rH^2-\frac{\kappa_{5}^{2}}{3}\,(\rho+\lambda)\Big{]}^2,
\ee where $\Phi$ is a solution to \be \label{Phi} \Phi +
2\alpha\,\Phi^2=\frac{\Lambda_{5}}{6}~, \ee which from
(\ref{GBIGL5}) gives two solutions for $\Phi$ \be \label{PhiSol}
\Phi_{1}=-\frac{1}{l^2}~, \,\,\,
\Phi_{2}=\frac{1}{l^2}-\frac{1}{2\alpha}~. \ee These solutions can
be combined together if we set  $\Phi = -\frac{\zeta}{l^2}$. Then,
defining \be \label{RedefGBIG} \tilde{l}^2 = -\frac{1}{\Phi} =
-\frac{l^2}{\zeta}, \ee the cubic equation \label{GBIGCubic} is
written as \be \label{GBIGCubic1} \kappa_{5}^{2}\, (\rho +
\lambda) - 3\,rH^2 = \frac{2}{\tilde{l}}\,
\sqrt{1+H^{2}\,\tilde{l}^{2}}\, \Big{(}3+\tilde{\beta}
\,(2H^{2}\,\tilde{l}^{2}-1)\Big{)}~, \ee where $\tilde{\beta} =
4\alpha/\tilde{l}^2$. Moreover, in order to recover the late time
GR cosmology we must also have $\tilde{l} \gg r$. We thus find for
this model four different regimes for the dynamical evolution of
the brane-universe:
\begin{itemize}
\item at high energy, the GB regime, when $H\tilde{l} \gg Hr \gg
\tilde{\beta}^{-1} \gg 1$ or $H\tilde{l} \gg \tilde{\beta}^{-1}
\gg Hr \gg 1$ \be \label{GBGBIGRegime} H^2\simeq \Big{[}
\frac{\kappa_{5}^2}{4\tilde{\beta}\, \tilde{l}^2}\, \rho \Big{]}
^{2/3}, \ee \item a $4D$ IND regime, when $\tilde{\beta}^{-1} \gg
H\tilde{l} \gg Hr \gg 1$ \footnote{We could have avoited the IND
regime and keep a GB behaviour if we had instead $H\tilde{l} \gg
\tilde{\beta}^{-1} \gg 1 \gg Hr$. However this would finally give
a pure GB behaviour which is not of any interest.} \be
\label{INDGBIGRegime} H^2\simeq \frac{\kappa_{4}^{2}\, \rho}{3}
\Big{(} 1 - \frac{2\sqrt{3}}{\kappa_{4}\, r\, \rho^{1/2}}
\Big{)}~, \ee \item an intermediate $5D$ RS regime, when
$\tilde{\beta}^{-1} \gg H\tilde{l} \gg 1 \gg Hr$ \be
\label{RSGBIGRegime} H^2\simeq\frac{\kappa_{5}^4}{36}\,
(\rho+\lambda)^{2}~, \ee \item the GR regime at low energy, when
$\tilde{\beta}^{-1} \gg 1 \gg H\tilde{l} \gg Hr$ \be
\label{GRGBIGRegime} H^2\simeq\frac{\kappa_{Pl}^2}{3}\, \rho~. \ee
\end{itemize}
The brane tension obeys the same fine-tuning relation as in the GB
model \be \label{FineTuneGBIG} \lambda = \frac{6}{\kappa^{5}\,
\tilde{l}}\,\Big{(}1-\frac{1}{3}\tilde{\beta} \Big{)}~, \ee and
the effective $4D$ Newton constant is \be \label{NewtonGBIG}
\kappa_{Pl}^2 =
\frac{\kappa_{5}^{2}}{\tilde{l}\,(1+\tilde{\beta})\,+r}~.
\ee

\subsection{Inflation and Reheating in the GBIG Brane-world Model}

In this cosmological model inflation can start during a GB, a $4D$
IND or a RS regime and it can end in the same or in a different
regime. Then the curvaton reheating follows. We can distinguish
the following cases:
\begin{itemize}
\item
\underline{Inflation occurs in the GB regime}

 We show in section
4.2 that inflation can not end in the GB region but it ends in the
RS regime and then the results of section 4.4 apply. If inflation
ends in the IND regime then the results of section 5.4 apply.

\item
\underline{Inflation occurs in the ING regime}

In this case the analysis of section 5.3 applies.

\item
\underline{Inflation occurs in the RS regime}

In this case the analysis of sections 5.3 and 5.4 applies.
\end{itemize}

\section{Conclusions and Discussion}

We studied the curvaton dynamics in brane-worlds. The curvaton
reheating mechanism was applied to various stages of the
cosmological evolution of the brane-world models. These models are
introducing unconventional correction terms to the Friedmann
equation of the standard cosmology. These terms have important
consequences to the inflationary dynamics. They make the inflation
easier because in most cases they act as friction terms. Also they
enhance the scalar and tensor perturbations generated during
inflation. However, these corrections terms are high energy
effects and as the energy density is decreasing, soon they
decouple from the cosmological dynamics. This leaves open the
possibility that the inflaton field survives without decay after
the end of inflation. Then, the curvaton field provides the
mechanism for the reheating of the universe.

We developed a general curvaton formalism appropriate to
brane-worlds. In all of the brane-world models there is a
transition from one dimensionality spacetime to another as the
energy density decreases. Thus the curvaton oscillates in one
dimensionality space and may decay in the same or at a different
dimensionality spacetime. Different dimensionalities  spacetimes
have different cosmological dynamics in brane-worlds and this is
translated to a system of constraints that the curvaton parameters
should respect.

We derived constrained relations which the curvaton parameters
should also satisfy in order to suppress short-wavelength
gravitational waves dominance over radiation. According to our
hypothesis, the inflaton field survives without decay after the
end of inflation. Then it enters a kinetic epoch until the
curvaton takes over and dominates the cosmological evolution. If
this epoch is long enough, there is a possibility of generation of
large amplitude gravitational waves. In brane-worlds there are
various interesting cases of competing or complementary effects.
For example, the inflaton field can enter without decay a
five-dimensional regime (induced gravity model). In this regime
the kinetic epoch does not last long because the correction terms
act as friction terms but at the same time the amplitude of
gravitational perturbations is enhanced. The analysis of various
such cases constrained further the curvaton parameters.

We investigated the possibility that density perturbations are
also generated by the currvaton field. Then we found that this is
not always possible. There are cases in which the curvaton
parameters are so constrained such as  curvaton fluctuations could
not generate density perturbations. Nevertheless, in all
brane-world models there are cases that density perturbations are
indeed generated by the curvaton field. We showed that in these
cases the upper bound of the fundamental $M_{5}$ mass decreases
considerably and it can take values much lower than the Planck
mass. Its final value depends on how much the inflaton
fluctuations are suppressed compared to curvaton fluctuations.

For the  Randall-Sundrum model we analysed in detail the case that
inflation and density perturbations are generated by the inflaton
field with a quadratic potential in the five-dimensional
spacetime, while reheating is done by the curvaton field in four
dimensions. This analysis indicates that inflation and primordial
density perturbations are pure five-dimensional high energy
effects, while curvaton dynamics decoupled from the high energy
regime can lead to low energy values of its parameters.

The analysis of the Gauss-Bonnet model, taking under consideration
the latest three-year WMAP results, showed that the most probable
case is the inflation to end in the five-dimensional
Randall-Sundrum regime reproducing in this way all the results for
the reheating of the Randall-Sundrum model. In the induced gravity
model we have the interesting possibility that inflation, density
perturbations and reheating  occur in the four-dimensional high
energy regime. Finally, in the combined Gauss-Bonnet and induced
gravity model with an appropriate choice of parameters we can
reproduce the results obtained for the other three brane-world
models.

The introduction of the curvaton field in the brane-worlds had
given a better understanding of the early time cosmological
dynamics of a brane-universe. This ``hybrid" inflaton-curvaton
model gives more freedom to constraint the parameters and study
interesting physically cases. However, a possible drawback for
such a successful ``hybrid" model is that the parameters should
satisfy a rather complex system of constraints.

\section{Acknowledgements}

We thank K. Dimopoulos, G. Lazarides, R. Maartens, M. Sami and D.
Wands for their valuable remarks, comments and suggestions. Work
supported by (EPEAEK II)-Pythagoras (co-funded by the European
Social Fund and National Resources).

\end{document}